\documentclass[a4paper,onecolumn,unpublished,11pt]{quantumarticle}
\pdfoutput=1

\usepackage[numbers,sort&compress]{natbib}
\usepackage[utf8]{inputenc}
\usepackage[T1]{fontenc}
\usepackage[english]{babel}
\usepackage{physics}
\usepackage{graphicx}
\usepackage{amsmath,amsfonts,amsthm,amssymb,mathtools}
\usepackage{enumitem}
\usepackage{thmtools}
\usepackage{hyperref}
\usepackage[capitalise,noabbrev]{cleveref}
\usepackage{float}
\usepackage[table]{xcolor}
\usepackage{bm}

\newtheorem{thm}{Theorem}

 \newtheorem{prop}[thm]{Proposition}
\newtheorem{cor}{Corollary}
\newtheorem{defn}{Definition}

\newtheorem{rem}{Remark}
\newtheorem{fact}{Fact}
\newtheorem{obs}{Observation}

\newcommand{\Sym}{\operatorname{Sym}}
\newcommand{\Herm}{\operatorname{Herm}}
\newcommand{\id}{\operatorname{id}}
\newcommand{\opket}[1]{\lvert #1\rangle\!\rangle}
\newcommand{\opbra}[1]{\langle\!\langle #1\rvert}

\begin{document}

\title{Approximate pushforward designs and image bounds on approximations}

\author{Jakub Czartowski}
\email{czartowj@tcd.ie}
\affiliation{School of Physics, Trinity College Dublin, Dublin 2, Ireland}
\affiliation{School of Physical and Mathematical Sciences, Nanyang Technological University, 21 Nanyang Link, 637361 Singapore, Republic of Singapore}
\affiliation{Faculty of Physics, Astronomy and Applied Computer Science, Jagiellonian University, 30-348 Krak\'ow, Poland}
\orcid{0000-0003-4062-833X}

\author{Adam Sawicki}
\email{a.sawicki@cft.edu.pl}
\affiliation{Center for Theoretical Physics, Polish Academy of Sciences, Aleja Lotnik\'ow 32/46, 02-668 Warszawa, Poland}
\affiliation{Guangdong Technion--Israel Institute of Technology, 241 Daxue Road, Jinping District, Shantou, Guangdong, China}

\author{Karol {\.Z}yczkowski}
\email{karol@cft.edu.pl}
\affiliation{Faculty of Physics, Astronomy and Applied Computer Science, Jagiellonian University, 30-348 Krak\'ow, Poland}
\affiliation{Center for Theoretical Physics, Polish Academy of Sciences, Aleja Lotnik\'ow 32/46, 02-668 Warszawa, Poland}

\date{August 17, 2026}

\begin{abstract}
We extend the framework of quantum pushforward designs to the approximate setting, in which moment operators agree only up to finite precision. We first formulate a general transfer theorem for maps that act linearly at the level of moments. This separates direct applications, controlled by standard Schatten-norm estimates, from refinements that use additional tensorial structure. Dephasing is shown to be contractive and therefore sends an approximate projective design to a simplex design without increasing its error. Ordinary partial-trace estimates give immediate bounds for mixed-state and channel designs. We then prove a sharp Schatten-norm bound for the partial trace restricted to the totally symmetric subspace. This replaces the full-environment norm coefficient by one governed by a symmetric-subspace dimension and yields asymptotically tighter estimates for both mixed-state and channel designs. Numerical simulations for induced mixed-state designs are consistent with the resulting hierarchy of bounds.
\end{abstract}

\maketitle

\section{Introduction}

Designs, also known as averaging sets or quadratures, provide powerful methods for evaluating polynomial functions over compact spaces with respect to a given measure. They have found diverse applications, from Gauss quadratures in numerical integration \cite{gauss1815methodus} to spherical designs \cite{Delsarte1977,SEYMOUR1984213}, which are used in fields such as electrostatics \cite{Simmonett2022}. More recently, complex projective designs and unitary designs have been applied to quantum state tomography and quantum channel tomography, respectively \cite{scott2006tight,scott2008optimizing}. These ideas have also been extended to continuous-variable systems \cite{iosue2024continuous}, noncompact spaces \cite{Markiewicz2023}, and constructions that combine designs on different spaces. For example, projective toric designs together with simplex designs produce complex projective designs of product form \cite{iosue2023projective}. Mapping designs from one space to another similarly leads to mixed-state designs and channel designs, collectively referred to as pushforward designs \cite{czartowski2019isoentangled,czartowski2024quantum}.

In parallel, approximate designs have become important when exact averaging is unavailable or unnecessary. Approximate complex projective designs are used to derandomize random measurements \cite{ambainis07}, while approximate unitary designs occur in the study of generic quantum speedups \cite{Hallgren2008superpolynomial}. Considerable attention has also been devoted to constructing approximate $t$-designs from random quantum circuits \cite{brandao2016local,Jian2023,leone2025non1clifford}. Low-order moment conditions, including fourth moments, additionally play an important role in criteria for verifying universality of quantum gate sets \cite{sawicki2017criteria,sawicki2022universality}.

\begin{figure}[h]
    \centering
    \includegraphics[width=0.75\linewidth]{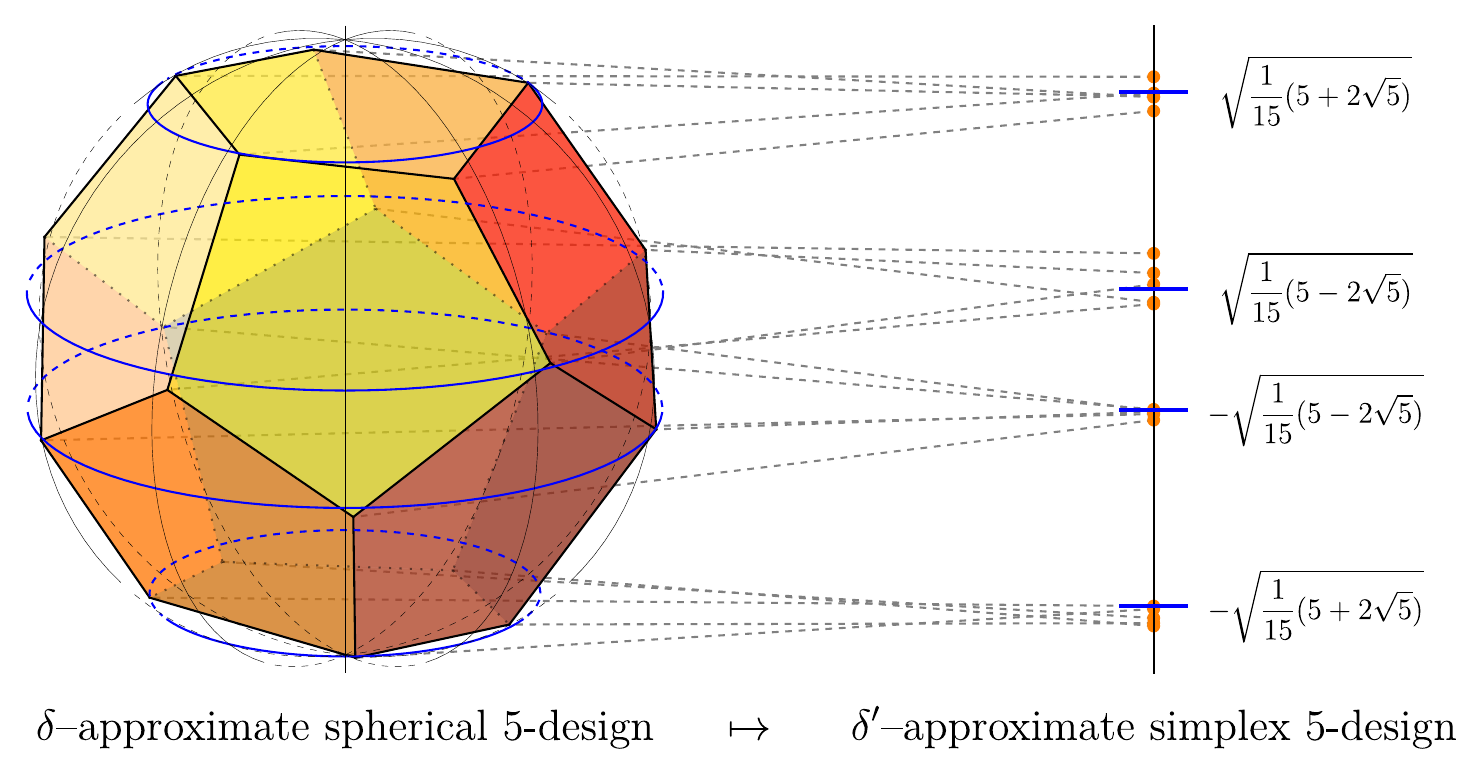}
    \caption{\textbf{Conceptual illustration of approximate pushforward designs:} The vertices of a regular dodecahedron form an exact spherical $5$-design, or equivalently a complex projective $5$-design on $\mathbb{C}P^1$ under the Bloch-sphere identification. Projecting this exact design onto a selected axis gives four exact projected values, shown by the blue levels at $ \pm\sqrt{\frac{1}{15}\bigl(5\pm 2\sqrt{5}\bigr)}$, and hence an exact degree-$5$ pushforward design on the one-dimensional simplex. After a small deformation of the dodecahedron, the $20$ vertices no longer project exactly onto these four values; instead, their images become $20$ slightly displaced points, shown in orange, clustered around the corresponding exact levels. The results of this work bound the induced output error $\delta'$ of this projected pushforward design in terms of the input error~$\delta$. }
    \label{fig:concept}
\end{figure}

In this work, we address the following general question, depicted conceptually in Fig.~\ref{fig:concept}. Given a $\delta_p$-approximate $t$-design on a space $X$, how large can the approximation error $\delta'_p$ become after the design is pushed forward to an image space $Y$? The central point is that the map between the underlying points need not itself be linear. What matters is whether the map induces a linear transformation between the corresponding moment operators. This distinction covers the quantum-information constructions considered here, including dephasing, partial trace, and the passage from bipartite unitaries to reduced quantum channels.

The paper is organized as follows. In Section~\ref{sec:prel}, we introduce balanced polynomials, moment operators, exact and approximate designs, and the norm estimates used later. In Section~\ref{sec:main}, we prove the main transfer theorem for pushforwards that are linear at the level of moment operators. Section~\ref{sec:direct_applications} collects the applications that follow directly from standard norm inequalities: simplex designs, mixed-state designs, and an unrefined channel-design bound. Section~\ref{sec:refined_applications} contains the more involved results. There we prove a sharp partial-trace estimate on the totally symmetric subspace and use it to improve the mixed-state and channel bounds. In Section~\ref{sec:num}, we test the mixed-state estimates numerically. Finally, Section~\ref{sec:conclusions} summarizes the results and discusses possible extensions.

\section{Preliminaries}
\label{sec:prel}

Throughout, $t$ is a positive integer. Let $\mathcal H_d\simeq\mathbb C^d$ be a $d$-dimensional Hilbert space and let $\mathbb P\mathcal H_d\equiv\mathbb CP^{d-1}$ denote the associated complex projective space. A point of $\mathbb P\mathcal H_d$ is a ray $[\psi]$; whenever convenient, we represent it by a normalized vector $\ket\psi$ satisfying $\braket\psi=1$. Two normalized representatives define the same ray precisely when they differ by a phase, $\ket\psi\sim e^{i\phi}\ket\psi$. Subscripts such as $A$ and $B$ label tensor factors, with $d_A=\dim\mathcal H_A$ and $d_B=\dim\mathcal H_B$.

We write $\mathcal L(\mathcal H)$ for the linear operators on $\mathcal H$ and
\begin{equation}
\mathcal B(\mathcal H)
=
\qty{\rho\in\mathcal L(\mathcal H):
\rho=\rho^\dagger,\ \rho\geq0,\ \Tr\rho=1}
\end{equation}
for the density operators. The unitary group on $\mathcal H_d$ is denoted by $\mathcal U(d)$. A quantum channel $\Phi:\mathcal L(\mathcal H)\to\mathcal L(\mathcal H')$ is a completely positive trace-preserving linear map. Every such map has a Kraus representation
\begin{equation}
\Phi(Q)=\sum_{i=1}^{r}K_iQK_i^\dagger,
\qquad
\sum_{i=1}^{r}K_i^\dagger K_i=\mathbb I.
\end{equation}
The least possible number $r$ of Kraus operators is the Choi rank, equivalently the dimension of a minimal Stinespring environment, and satisfies $r\leq\dim(\mathcal H)\dim(\mathcal H')$.

For a finite-dimensional Hilbert space $V$, let $\Sym_t(V)\subset V^{\otimes t}$ be the totally symmetric subspace. If $P_V(\sigma)$ denotes the unitary that permutes the $t$ tensor factors according to $\sigma\in S_t$, then the orthogonal projector onto this subspace is
\begin{equation}
\label{eq:symmetric_projector}
\Pi_{V,t}
=
\frac{1}{t!}\sum_{\sigma\in S_t}P_V(\sigma).
\end{equation}
Its dimension is
\begin{equation}
\label{eq:symmetric_dimension}
D_{V,t}:=\dim\Sym_t(V)=\binom{\dim V+t-1}{t}.
\end{equation}

\subsection{Balanced polynomials and moment operators}
\label{sec:balanced}

Let $V$ be a finite-dimensional complex Hilbert space. A balanced polynomial of bidegree $(t,t)$ is a polynomial $g_t:V\to\mathbb C$ that is homogeneous of degree $t$ in the coordinates of $x$ and of degree $t$ in their complex conjugates. Thus
\begin{equation}
g_t(cx)=|c|^{2t}g_t(x),
\qquad c\in\mathbb C.
\end{equation}
In particular, $g_t(e^{i\phi}x)=g_t(x)$, so its restriction to normalized vectors defines a function on projective space.

Because ordinary coordinates commute, the coefficient tensor of $g_t$ can be symmetrized separately in its holomorphic and antiholomorphic indices. The symmetrized holomorphic monomials form a basis of $\Sym_t(V)$, and the antiholomorphic monomials form the conjugate basis. Thus the symmetrized coefficients are precisely the matrix elements of a unique operator
\begin{equation}
G_t\in\mathcal L\qty(\Sym_t(V))
\end{equation}
such that
\begin{equation}
\label{eq:balanced_operator_form}
g_t(x)
=
\mel{x^{\otimes t}}{G_t}{x^{\otimes t}}
=
\Tr\qty[G_t\qty(\ket{x}\bra{x})^{\otimes t}].
\end{equation}
If $G_t$ is regarded as an operator on all of $V^{\otimes t}$, the convention $G_t=\Pi_{V,t}G_t\Pi_{V,t}$ is essential for uniqueness: components acting outside the symmetric subspace do not contribute to Eq.~\eqref{eq:balanced_operator_form}. Complex conjugation interchanges the holomorphic and antiholomorphic coefficient indices. Hence $g_t(x)$ is real for every $x$ precisely when the coefficient matrix satisfies~$G_t=G_t^\dagger$.

For a real Hilbert space $V$, the corresponding object is a homogeneous polynomial of degree $t$. It has a unique symmetric coefficient tensor $G_t\in\Sym_t(V)$ for which
\begin{equation}
\label{eq:real_polynomial_form}
g_t(x)=\langle G_t,x^{\otimes t}\rangle.
\end{equation}
When the points $x$ are themselves Hermitian matrices, the Hilbert--Schmidt pairing allows the same expression to be written as $g_t(x)=\Tr(O_tx^{\otimes t})$ for a suitable coefficient operator~$O_t$.

The concrete moment operators used in this paper are as follows.
\begin{itemize}[leftmargin=*]
\item \textbf{Pure states.} For a probability measure $\lambda$ on $\mathbb P\mathcal H_d$,
\begin{equation}
T^{\mathrm{proj}}_{\lambda,t}
=
\int_{\mathbb P\mathcal H_d}
\qty(\ket\psi\bra\psi)^{\otimes t}\,d\lambda(\psi).
\end{equation}
Every integrand, and hence the moment operator, is supported on $\Sym_t(\mathcal H_d)$.

\item \textbf{Mixed states.} For a probability measure $\lambda$ on $\mathcal B(\mathcal H_d)$,
\begin{equation}
T^{\mathrm{mix}}_{\lambda,t}
=
\int_{\mathcal B(\mathcal H_d)}\rho^{\otimes t}\,d\lambda(\rho).
\end{equation}
Here $\mathcal B(\mathcal H_d)$ is viewed as a subset of the real vector space $\Herm(\mathcal H_d)$.

\item \textbf{Unitaries.} The standard unitary moment of a probability measure $\lambda$ on $\mathcal U(d)$ is
\begin{equation}
\label{eq:standard_unitary_moment}
T^{\mathrm U}_{\lambda,t}
=
\int_{\mathcal U(d)}
U^{\otimes t}\otimes\overline U^{\otimes t}\,d\lambda(U).
\end{equation}
This convention directly records balanced polynomials in the matrix entries of $U$. A normalized rank-one, or Choi-vector, convention will be introduced in Section~\ref{subsec:direct_channel}. The two conventions are related by realignment, but they are not isometric in every Schatten norm; this point is treated explicitly in Remark~\ref{rem:realignment_conversion}.
\end{itemize}

To avoid cumbersome superscripts when the underlying space is clear, we write all three kinds of moment operators as $T_{\lambda,t}$.

\subsection{Exact and approximate designs}

\begin{defn}[Exact $t$-design]
\label{def:gen}
Let $X$ be a measurable space, let $\mathcal K_X$ be a finite-dimensional Hilbert space, and let
\begin{equation}
M_{X,t}:X\longrightarrow\mathcal L(\mathcal K_X)
\end{equation}
be a bounded measurable operator-valued map, called the \emph{moment kernel}. For every probability measure $\lambda$ on $X$, define
\begin{equation}
T^X_{\lambda,t}
:=
\int_X M_{X,t}(x)\,d\lambda(x).
\end{equation}
Fix a target probability measure $\mu_X$ on $X$. A probability measure $\nu_X$ on $X$ is an exact $t$-design relative to $\mu_X$ and the chosen moment kernel if
\begin{equation}
T^X_{\nu_X,t}=T^X_{\mu_X,t}.
\end{equation}
Equivalently, $\mu_X$ and $\nu_X$ give the same average for every test function
\begin{equation}
x\longmapsto
\Tr\qty[G^\dagger M_{X,t}(x)],
\qquad
G\in\mathcal L(\mathcal K_X).
\end{equation}
For the concrete kernels introduced above, these test functions are precisely the polynomials encoded by the corresponding moment operators.
\end{defn}

\begin{rem}[Moment kernels]
The moment kernel is the fixed operator-valued function whose average defines the moment operator; it is not a probability transition kernel. Measurability means that
\begin{equation}
x\longmapsto\mel{u}{M_{X,t}(x)}{v}
\end{equation}
is measurable for every $u,v\in\mathcal K_X$. Since $\mathcal K_X$ is finite-dimensional, this is equivalent to measurability of all matrix entries in any orthonormal basis. Boundedness means that
\begin{equation}
\sup_{x\in X}\norm{M_{X,t}(x)}_\infty<\infty.
\end{equation}
These assumptions ensure that the operator-valued integral in Definition~\ref{def:gen} is well defined. In the applications considered below, the spaces are compact and the moment kernels are continuous, so both conditions hold automatically.
\end{rem}

The definition for a general moment kernel does not by itself relate different orders. For the normalized state kernels and the balanced unitary kernels used in this paper, however, there is an explicit linear map that sends the moment of order $t$ to the moment of every lower order $s<t$. Consequently, equality of the $t$-th moments implies equality of the $s$-th moments.

\noindent For pure and mixed states, let
\begin{equation}
\mathcal C^{\mathrm{state}}_{t\to s}
:=
\Tr_{s+1,\ldots,t}
\end{equation}
denote the partial trace over the last $t-s$ tensor factors. If $\rho$ is a density operator, then
\begin{align}
\mathcal C^{\mathrm{state}}_{t\to s}
\qty(\rho^{\otimes t})
&=
\rho^{\otimes s}.
\end{align}
Here one may take
$\rho=\ket\psi\bra\psi$ in the pure-state case. Since partial trace is linear, it commutes with integration, and hence
\begin{align}
\mathcal C^{\mathrm{state}}_{t\to s}
\qty(T_{\lambda,t})
&=
\int
\mathcal C^{\mathrm{state}}_{t\to s}
\qty(\rho^{\otimes t})\,d\lambda(\rho)\\
&=
T_{\lambda,s}.
\end{align}
Therefore,
\begin{equation}
T_{\nu,t}=T_{\mu,t}
\quad\Longrightarrow\quad
T_{\nu,s}
=
\mathcal C^{\mathrm{state}}_{t\to s}(T_{\nu,t})
=
\mathcal C^{\mathrm{state}}_{t\to s}(T_{\mu,t})
=
T_{\mu,s}.
\end{equation}
Thus every exact projective or mixed-state $t$-design is also an exact design of every order $s<t$.
\noindent For unitary moments, recall that
\begin{equation*}
T^{\mathrm U}_{\lambda,t}
=
\int_{\mathcal U(d)}
U^{\otimes t}\otimes\overline U^{\otimes t}\,
d\lambda(U).
\end{equation*}
The tensor factors in this expression are initially ordered as
\begin{equation}
U_1\otimes\cdots\otimes U_t
\otimes
\overline U_1\otimes\cdots\otimes\overline U_t.
\end{equation}
For every positive integer $r$, let $W_r$ be the fixed unitary permutation of tensor factors defined by
\begin{equation}
W_r
\qty(
\ket{x_1}\otimes\cdots\otimes\ket{x_r}
\otimes
\ket{y_1}\otimes\cdots\otimes\ket{y_r}
)
=
\bigotimes_{j=1}^r
\qty(\ket{x_j}\otimes\ket{y_j}).
\end{equation}
This permutation is independent of $U$ and merely places each factor $U_j$ next to the corresponding factor $\overline U_j$. Consequently,
\begin{equation}
W_r
\qty(
U^{\otimes r}\otimes\overline U^{\otimes r}
)
W_r^\dagger
=
\qty(U\otimes\overline U)^{\otimes r}.
\end{equation}

\noindent To remove one paired factor, introduce the normalized maximally entangled vector
\begin{equation}
\ket{\Omega_d}
=
\frac{1}{\sqrt d}
\sum_{j=1}^d
\ket j\otimes\ket j.
\end{equation}
Its purpose is to define a fixed linear contraction under which every pair
$U\otimes\overline U$ evaluates to one. Indeed,
\begin{align}
\qty(U\otimes\overline U)\ket{\Omega_d}
&=
\frac{1}{\sqrt d}
\sum_{j=1}^d
U\ket j\otimes\overline U\ket j\\
&=
\frac{1}{\sqrt d}
\sum_{a,b=1}^d
\qty(
\sum_{j=1}^d
U_{aj}\overline{U_{bj}}
)
\ket a\otimes\ket b\\
&=
\frac{1}{\sqrt d}
\sum_{a,b=1}^d
(UU^\dagger)_{ab}
\ket a\otimes\ket b\\
&=
\frac{1}{\sqrt d}
\sum_{a=1}^d
\ket a\otimes\ket a\\
&=
\ket{\Omega_d}.
\end{align}
Hence
\begin{equation}
\mel{\Omega_d}{U\otimes\overline U}{\Omega_d}=1.
\end{equation}

\noindent We may now define a linear reduction map from order $t$ to order $s$ by
\begin{equation}
\mathcal C^{\mathrm U}_{t\to s}(A)
=
W_s^\dagger
\qty(
\mathbb I\otimes
\bra{\Omega_d}^{\otimes(t-s)}
)
W_t A W_t^\dagger
\qty(
\mathbb I\otimes
\ket{\Omega_d}^{\otimes(t-s)}
)
W_s,
\end{equation}
where $\mathbb I$ acts on the first $s$ paired tensor factors. Applying this map to a unitary-moment integrand gives
\begin{align}
\mathcal C^{\mathrm U}_{t\to s}
\qty(
U^{\otimes t}\otimes\overline U^{\otimes t}
)
&=
W_s^\dagger
\qty(U\otimes\overline U)^{\otimes s}
W_s
\qty(
\mel{\Omega_d}{U\otimes\overline U}{\Omega_d}
)^{t-s}\\
&=
W_s^\dagger
\qty(U\otimes\overline U)^{\otimes s}
W_s\\
&=
U^{\otimes s}\otimes\overline U^{\otimes s}.
\end{align}
By linearity,
\begin{align}
\mathcal C^{\mathrm U}_{t\to s}
\qty(T^{\mathrm U}_{\lambda,t})
&=
\int_{\mathcal U(d)}
\mathcal C^{\mathrm U}_{t\to s}
\qty(
U^{\otimes t}\otimes\overline U^{\otimes t}
)
\,d\lambda(U)\\
&=
\int_{\mathcal U(d)}
U^{\otimes s}\otimes\overline U^{\otimes s}
\,d\lambda(U)\\
&=
T^{\mathrm U}_{\lambda,s}.
\end{align}
It follows that
\begin{equation}
T^{\mathrm U}_{\nu,t}
=
T^{\mathrm U}_{\mu,t}
\quad\Longrightarrow\quad
T^{\mathrm U}_{\nu,s}
=
\mathcal C^{\mathrm U}_{t\to s}
\qty(T^{\mathrm U}_{\nu,t})
=
\mathcal C^{\mathrm U}_{t\to s}
\qty(T^{\mathrm U}_{\mu,t})
=
T^{\mathrm U}_{\mu,s}.
\end{equation}
Thus equality of the balanced unitary moment of order $t$ implies equality at every balanced bidegree $(s,s)$ with $s<t$. This argument does not impose any condition on unbalanced monomials, whose degrees in the entries of $U$ and $\overline U$ are different.
\medskip

The target measure is typically fixed by the geometry: the Fubini--Study measure on $\mathbb P\mathcal H_d$, the Haar measure on $\mathcal U(d)$, or an induced measure on mixed states. If $\nu_X$ is a finite atomic measure, its support is
\begin{equation}
S_{\nu_X}=\qty{x\in X:\nu_X(\{x\})>0}.
\end{equation}
The number $w_x=\nu_X(\{x\})$ is the weight of $x$. The design is uniform when all these weights are equal. This formulation includes quadratures \cite{gauss1815methodus}, spherical designs \cite{Delsarte1977}, averaging sets \cite{SEYMOUR1984213}, complex projective designs \cite{Neumaier1981,hoggar1982,Hoggar1984,BannaiHoggar1985,Hoggar1989,Hoggar1992}, and unitary designs \cite{dankert2009exact}.
\medskip

For $Q\in\mathcal L(\mathcal K_1,\mathcal K_2)$ and $1\leq p<\infty$, the Schatten $p$-norm is
\begin{equation}
\norm{Q}_p
=
\qty(\Tr|Q|^p)^{1/p}
=
\qty(\sum_i s_i(Q)^p)^{1/p},
\end{equation}
where $s_i(Q)$ are the singular values. At the endpoint,
\begin{equation}
\norm{Q}_\infty=\max_i s_i(Q).
\end{equation}
The cases $p=1,2,\infty$ are the trace, Hilbert--Schmidt, and spectral norms, respectively.

\begin{defn}[$\delta_p$-approximate $t$-design]
\label{def:approximate_design}
For $1\leq p\leq\infty$, define the actual moment error by
\begin{equation}
\varepsilon_{p,t}(\mu_X,\nu_X)
:=
\norm{T^X_{\mu_X,t}-T^X_{\nu_X,t}}_p.
\end{equation}
The measure $\nu_X$ is a $\delta_p$-approximate $t$-design relative to $\mu_X$ when
\begin{equation}
\varepsilon_{p,t}(\mu_X,\nu_X)\leq\delta_p.
\end{equation}
\end{defn}

The definition is representation-dependent: an invertible rearrangement of tensor indices preserves exact moment equality, but need not preserve a Schatten-$p$ approximation error. We will therefore state explicitly which unitary-moment convention is used in the channel applications.

\subsection{Auxiliary norm bounds}
\label{sec:aux_norm_bounds}

We collect the norm estimates used below. For a linear map
\begin{equation}
\mathcal L:\mathcal L(\mathcal K_1)\to\mathcal L(\mathcal K_2),
\end{equation}
its induced Schatten norm is
\begin{equation}
\label{eq:induced_norm}
\norm{\mathcal L}_{p\to q}
:=
\sup_{Q\neq0}\frac{\norm{\mathcal L(Q)}_q}{\norm{Q}_p}.
\end{equation}
This notation distinguishes a superoperator norm from the spectral norm $\norm{F}_\infty$ of an ordinary matrix $F$.

The ideal property of Schatten norms states that
\begin{equation}
\label{eq:schatten_ideal}
\norm{AQB}_p
\leq
\norm{A}_\infty\norm{Q}_p\norm{B}_\infty.
\end{equation}
We also use the standard comparison
\begin{equation}
\label{eq:schatten_comparison}
\norm{Q}_q
\leq
\operatorname{rank}(Q)^{\max\{0,1/q-1/p\}}\norm{Q}_p,
\qquad 1\leq p,q\leq\infty.
\end{equation}

\begin{fact}[Partial-trace bound]
\label{fact:partial_trace_bound}
Let $\Tr_B:\mathcal L(\mathcal H_A\otimes\mathcal H_B)\to\mathcal L(\mathcal H_A)$, where $d_B=\dim\mathcal H_B$. Then, for $1\leq p\leq\infty$ \cite[Proposition~1]{rastegin2012relations},
\begin{equation}
\norm{\Tr_B Q}_p
\leq
d_B^{1-1/p}\norm{Q}_p,
\end{equation}
where the exponent is interpreted as $1$ for $p=\infty$.
\end{fact}

In particular,
\begin{equation}
\norm{\Tr_B Q}_1\leq\norm{Q}_1,
\qquad
\norm{\Tr_B Q}_2\leq\sqrt{d_B}\norm{Q}_2,
\qquad
\norm{\Tr_B Q}_\infty\leq d_B\norm{Q}_\infty.
\end{equation}

\begin{fact}[Channel norm bound]
\label{fact:channel_lipschitz_bound}
If $\Phi$ is a quantum channel of Choi rank $r$, then \cite[Corollary~1]{rastegin2012relations}
\begin{equation}
\norm{\Phi(Q)}_p
\leq
r^{1-1/p}\norm{Q}_p,
\qquad 1\leq p\leq\infty.
\end{equation}
\end{fact}

\noindent Finally, let $\{P_j\}_{j=0}^{m-1}$ be mutually orthogonal projections satisfying
\begin{equation}
P_jP_k=\delta_{jk}P_j,
\qquad
\sum_{j=0}^{m-1}P_j=\mathbb I.
\end{equation}
The map
\begin{equation}
\mathcal E(Q)
=
\sum_{j=0}^{m-1}P_jQP_j
\end{equation}
is conventionally called the \emph{pinching map} associated with the projections $\{P_j\}$. It removes the off-diagonal blocks $P_jQP_k$ with $j\neq k$ and retains only the diagonal blocks $P_jQP_j$.

We now show directly that pinching is contractive for every unitarily invariant norm. Let
\begin{equation}
\omega=e^{2\pi i/m},
\qquad
V=\sum_{j=0}^{m-1}\omega^jP_j.
\end{equation}
The operator $V$ is unitary because
\begin{equation}
VV^\dagger
=
\sum_{j,k=0}^{m-1}
\omega^{j-k}P_jP_k
=
\sum_{j=0}^{m-1}P_j
=
\mathbb I.
\end{equation}
For every integer $k$,
\begin{equation}
V^kQ(V^\dagger)^k
=
\sum_{j,\ell=0}^{m-1}
\omega^{k(j-\ell)}P_jQP_\ell.
\end{equation}
Averaging over $k=0,\ldots,m-1$ and using
\begin{equation}
\frac{1}{m}
\sum_{k=0}^{m-1}
\omega^{k(j-\ell)}
=
\delta_{j\ell}
\end{equation}
gives
\begin{align}
\frac{1}{m}
\sum_{k=0}^{m-1}
V^kQ(V^\dagger)^k
&=
\sum_{j,\ell=0}^{m-1}
\left(
\frac{1}{m}
\sum_{k=0}^{m-1}
\omega^{k(j-\ell)}
\right)
P_jQP_\ell\\
&=
\sum_{j=0}^{m-1}P_jQP_j\\
&=
\mathcal E(Q).
\end{align}
Thus the roots-of-unity average is simply a convenient way to cancel all off-diagonal blocks of $Q$.

If $\norm{\cdot}$ is any unitarily invariant norm, then the triangle inequality and unitary invariance imply
\begin{align}
\norm{\mathcal E(Q)}
&=
\norm{
\frac{1}{m}
\sum_{k=0}^{m-1}
V^kQ(V^\dagger)^k
}\\
&\leq
\frac{1}{m}
\sum_{k=0}^{m-1}
\norm{V^kQ(V^\dagger)^k}\\
&=
\frac{1}{m}
\sum_{k=0}^{m-1}
\norm{Q}\\
&=
\norm{Q}.
\end{align}
In particular, for every Schatten norm,
\begin{equation}
\label{eq:pinching_contraction}
\norm{\mathcal E(Q)}_p
\leq
\norm{Q}_p.
\end{equation}
Hence
\begin{equation}
\norm{\mathcal E}_{p\to p}\leq1.
\end{equation}
Conversely, if $P_j\neq0$, then $\mathcal E(P_j)=P_j$, and therefore
\begin{equation}
\norm{\mathcal E}_{p\to p}=1.
\end{equation}
\section{The main theorem}
\label{sec:main}

Let $X$ and $Y$ be compact Hausdorff spaces with their Borel $\sigma$-algebras. Let $\mathcal K_X$ and $\mathcal K_Y$ be finite-dimensional Hilbert spaces, and equip $X$ and $Y$ with continuous operator-valued moment kernels
\begin{equation}
M_{X,t}:X\to\mathcal L(\mathcal K_X),
\qquad
M_{Y,t}:Y\to\mathcal L(\mathcal K_Y).
\end{equation}
Continuity on a compact space makes both kernels bounded. Thus a probability measure $\lambda$ on $X$ has the well-defined moment operator
\begin{equation}
T^X_{\lambda,t}=\int_XM_{X,t}(x)\,d\lambda(x),
\end{equation}
and analogously on $Y$. Let $F:X\to Y$ be Borel measurable. Its pushforward $F_*\lambda$ is defined by
\begin{equation}
(F_*\lambda)(B)=\lambda\qty(F^{-1}(B))
\end{equation}
for every measurable set $B\subseteq Y$.

The useful condition is not that $F$ itself be linear, but that its action be linear after passing to the chosen moments.

\begin{defn}[Moment-linear pushforward]
\label{def:moment_linear}
The map $F:X\to Y$ is \emph{moment-linear at order $t$} if there is a linear map
\begin{equation}
\mathcal F_t:\mathcal L(\mathcal K_X)\longrightarrow\mathcal L(\mathcal K_Y)
\end{equation}
such that
\begin{equation}
\label{eq:moment_intertwining}
M_{Y,t}(F(x))
=
\mathcal F_t\qty(M_{X,t}(x))
\end{equation}
for every $x\in X$.
\end{defn}

\begin{thm}[Transfer of approximate designs]
\label{thm:moment_linear_pushforward}
Let $F:X\to Y$ be moment-linear at order $t$, with induced map $\mathcal F_t$. Let $\mu_X$ be a target probability measure on $X$, let $\nu_X$ be another probability measure, and set
\begin{equation}
\mu_Y=F_*\mu_X,
\qquad
\nu_Y=F_*\nu_X.
\end{equation}
Then, for $1\leq p,q\leq\infty$,
\begin{equation}
\label{eq:main_transfer_bound}
\norm{T^Y_{\mu_Y,t}-T^Y_{\nu_Y,t}}_q
\leq
\norm{\mathcal F_t}_{p\to q}
\norm{T^X_{\mu_X,t}-T^X_{\nu_X,t}}_p.
\end{equation}
Consequently, a $\delta_p$-approximate $t$-design on $X$ pushes forward to a $\delta'_q$-approximate $t$-design on $Y$ with
\begin{equation}
\delta'_q\leq\norm{\mathcal F_t}_{p\to q}\delta_p.
\end{equation}
\end{thm}

\begin{proof}
The kernels are bounded and take values in finite-dimensional operator spaces, so the following integrals are well defined. The change-of-variables identity for pushforward measures and the intertwining relation~\eqref{eq:moment_intertwining} give
\begin{align}
T^Y_{F_*\lambda,t}
&=
\int_YM_{Y,t}(y)\,d(F_*\lambda)(y)\\
&=
\int_XM_{Y,t}(F(x))\,d\lambda(x)\\
&=
\int_X\mathcal F_t\qty(M_{X,t}(x))\,d\lambda(x)\\
&=
\mathcal F_t\qty(T^X_{\lambda,t}).
\end{align}
Applying this identity to $\mu_X$ and $\nu_X$ and using linearity yields
\begin{equation}
T^Y_{\mu_Y,t}-T^Y_{\nu_Y,t}
=
\mathcal F_t\qty(T^X_{\mu_X,t}-T^X_{\nu_X,t}).
\end{equation}
The definition~\eqref{eq:induced_norm} now proves inequality~\eqref{eq:main_transfer_bound}.
\end{proof}

Theorem~\ref{thm:moment_linear_pushforward} also shows immediately that exact designs are preserved: when the input moment difference is zero, so is its image. The approximate problem consists of estimating the induced norm of $\mathcal F_t$, possibly after restricting its domain to the support on which the relevant moment differences live.

\begin{cor}[Linear maps between complex Hilbert spaces]
\label{cor:complex_linear_pushforward}
Let $V$ and $W$ be finite-dimensional complex Hilbert spaces, let $X\subset V$ be compact, let $F:V\to W$ be linear, and let $\mu$ and $\nu$ be probability measures on $X$. Equip $X$ and $F(X)$ with the projective-type kernels
\begin{equation}
M_{X,t}(x)=\qty(\ket x\bra x)^{\otimes t},
\qquad
M_{F(X),t}(y)=\qty(\ket y\bra y)^{\otimes t}.
\end{equation}
The induced moment map is
\begin{equation}
\mathcal F_t(Q)=F^{\otimes t}Q(F^\dagger)^{\otimes t}.
\end{equation}
Therefore, for every $1\leq p\leq\infty$,
\begin{equation}
\norm{T_{F_*\mu,t}-T_{F_*\nu,t}}_p
\leq
\norm{F}_\infty^{2t}\norm{T_{\mu,t}-T_{\nu,t}}_p.
\end{equation}
\end{cor}

\begin{proof}
For every $x\in V$,
\begin{equation}
\qty(\ket{Fx}\bra{Fx})^{\otimes t}
=
F^{\otimes t}\qty(\ket x\bra x)^{\otimes t}(F^\dagger)^{\otimes t}.
\end{equation}
The result follows from Theorem~\ref{thm:moment_linear_pushforward}, the ideal property~\eqref{eq:schatten_ideal}, and
\begin{equation}
\norm{F^{\otimes t}}_\infty=\norm{F}_\infty^t.
\end{equation}
\end{proof}

\begin{cor}[Linear maps on operator-valued points]
\label{cor:operator_pushforward}
Let $X\subset\Herm(\mathcal H)$ be compact, let $\Phi:\mathcal L(\mathcal H)\to\mathcal L(\mathcal K)$ be linear and Hermiticity-preserving, and let $\mu$, $\nu$, and $\lambda$ be probability measures on $X$. Equip $X$ and $\Phi(X)$ with the kernels
\begin{equation}
M_{X,t}(Q)=Q^{\otimes t},
\qquad
M_{\Phi(X),t}(R)=R^{\otimes t}.
\end{equation}
Then
\begin{equation}
T_{\Phi_*\lambda,t}=\Phi^{\otimes t}(T_{\lambda,t})
\end{equation}
and hence, for every $1\leq p\leq\infty$,
\begin{equation}
\norm{T_{\Phi_*\mu,t}-T_{\Phi_*\nu,t}}_p
\leq
\norm{\Phi^{\otimes t}}_{p\to p}
\norm{T_{\mu,t}-T_{\nu,t}}_p.
\end{equation}
If $\Phi$ is a channel of Choi rank $r$, then
\begin{equation} \label{eq:tensor_channel_bound}
\norm{\Phi^{\otimes t}}_{p\to p}
\leq
r^{t(1-1/p)}.
\end{equation}
\end{cor}

\begin{proof}
For every $Q\in X$, tensoring the identity $\Phi(Q)$ with itself $t$ times gives
\begin{equation}
M_{\Phi(X),t}(\Phi(Q))
=
\Phi(Q)^{\otimes t}
=
\Phi^{\otimes t}\qty(Q^{\otimes t}).
\end{equation}
Thus the moment-level map is $\mathcal F_t=\Phi^{\otimes t}$. Integrating proves the first identity, and Theorem~\ref{thm:moment_linear_pushforward} gives the norm inequality. If $\Phi$ is a channel of Choi rank $r$, Eq.~\eqref{eq:tensor_channel_bound} gives the final estimate.
\end{proof}

The appearance of $\Phi^{\otimes t}$ is important. For a general superoperator, one must not replace $\norm{\Phi^{\otimes t}}_{p\to p}$ by $\norm{\Phi}_{p\to p}^t$ without a separate multiplicativity argument. In the applications below, the required tensor-power bounds follow directly from the structure of a pinching or a partial trace.

\section{Direct applications of the main theorem}
\label{sec:direct_applications}

We first collect the cases in which Theorem~\ref{thm:moment_linear_pushforward} can be combined directly with a standard norm inequality. No restriction beyond the ambient operator space is used in this section. The more delicate improvements obtained from symmetric input support are postponed to Section~\ref{sec:refined_applications}.

\subsection{Simplex designs from complex projective designs}
\label{subsec:simplex_designs}

Fix an orthonormal basis $\{\ket j\}_{j=1}^d$ of $\mathcal H_d$. The $(d-1)$-dimensional probability simplex is
\begin{equation}
\Delta_{d-1}
:=
\left\{
\mathbf p=(p_1,\ldots,p_d)\in\mathbb R^d:
p_j\geq0\ \text{for all }j,
\quad
\sum_{j=1}^d p_j=1
\right\}.
\end{equation}
Every normalized pure state $\ket\psi\in\mathcal H_d$ determines the probability vector
\begin{equation}
\mathbf p_\psi
=
\qty(
|\braket{1}{\psi}|^2,
\ldots,
|\braket{d}{\psi}|^2
)
\in\Delta_{d-1}.
\end{equation}
For $\mathbf p=(p_1,\ldots,p_d)\in\Delta_{d-1}$, write
\begin{equation}
\rho(\mathbf p):=\sum_{j=1}^dp_j\ket j\bra j
\end{equation}
and use the simplex moment kernel
\begin{equation}
\label{eq:simplex_moment_kernel}
M_{\Delta,t}(\mathbf p)=\rho(\mathbf p)^{\otimes t},
\qquad
T^\Delta_{\lambda,t}=\int_{\Delta_{d-1}}M_{\Delta,t}(\mathbf p)\,d\lambda(\mathbf p).
\end{equation}
The matrix entries of this kernel are the homogeneous monomials of total degree $t$ in the coordinates $p_j$. Because $\sum_jp_j=1$, a monomial $m$ of degree $s<t$ satisfies
\begin{equation}
m(\mathbf p)=m(\mathbf p)\qty(\sum_jp_j)^{t-s},
\end{equation}
whose expansion is a linear combination of degree-$t$ monomials. Thus this convention also controls all lower simplex degrees.
\medskip

\noindent The map from pure states to probability vectors can be represented by the completely dephasing channel
\begin{equation}
\label{eq:dephasing_channel}
\mathcal D:\mathcal L(\mathcal H_d)\longrightarrow\mathcal L(\mathcal H_d),
\qquad
\mathcal D(Q)
=
\sum_{j=1}^d
\ket j\bra jQ\ket j\bra j.
\end{equation}
Indeed,
\begin{align}
\mathcal D\qty(\ket\psi\bra\psi)
&=
\sum_{j=1}^d
|\braket{j}{\psi}|^2\ket j\bra j\\
&=
\rho(\mathbf p_\psi).
\end{align}
Thus, after identifying diagonal density operators with their diagonal probability vectors, the map
\begin{equation}
[\psi]\longmapsto\mathbf p_\psi
\end{equation}
is precisely the restriction of $\mathcal D$ to pure states.

Let $\mu_{\mathcal H}$ be the Fubini--Study probability measure on
$\mathbb P\mathcal H_d$, and let $\nu_{\mathcal H}$ be another probability measure on the same space. Define their pushforwards to the simplex by
\begin{equation}
\mu_\Delta
:=
(\mathbf p_{\bullet})_*\mu_{\mathcal H},
\qquad
\nu_\Delta
:=
(\mathbf p_{\bullet})_*\nu_{\mathcal H},
\end{equation}
where $\mathbf p_{\bullet}:[\psi]\mapsto\mathbf p_\psi$. The reference measure $\mu_\Delta$ is the Dirichlet distribution with parameter vector $(1,\ldots,1)$, equivalently normalized uniform Lebesgue measure on $\Delta_{d-1}$. This identifies the target measure for simplex designs. The error estimate below uses the representation by $\mathcal D$, together with the fact that dephasing is contractive in every Schatten norm.

Note that an example of such a construction is sketched in Fig.~\ref{fig:concept}: 
A regular dodecahedron forms a projective $5$-design in 
${\cal H}_2$, while its projection onto rotation axis due to decoherence with respect to the  corresponding basis yields
four points on the interval which form a
simplex $5$-design. It is possible 
to show, how the accuracy of the 
initial approximate projective design influences the precision of the 
 simplex design induced by the dephasing channel.

\begin{thm}[Simplex designs from projective designs]
\label{thm:simplex_from_projective}
If $\nu_{\mathcal H}$ is a $\delta_p$-approximate complex projective $t$-design relative to the Fubini--Study measure $\mu_{\mathcal H}$, then its image $\nu_\Delta$ is a $\delta'_p$-approximate simplex $t$-design satisfying
\begin{equation}
\label{eq:simplex_contraction}
\delta'_p
=
\norm{T_{\mu_\Delta,t}-T_{\nu_\Delta,t}}_p
\leq
\norm{T_{\mu_{\mathcal H},t}-T_{\nu_{\mathcal H},t}}_p
\leq
\delta_p.
\end{equation}
\end{thm}

\begin{proof}
For every normalized $\ket\psi$ one has,
\begin{equation}
M_{\Delta,t}(\mathbf p_\psi)
=
\mathcal D^{\otimes t}\qty[
\qty(\ket\psi\bra\psi)^{\otimes t}].
\end{equation}
Thus, after diagonal density matrices are identified with probability vectors, the pushforward moments satisfy
\begin{equation}
T_{\nu_\Delta,t}
=
\mathcal D^{\otimes t}\qty(T_{\nu_{\mathcal H},t}),
\qquad
T_{\mu_\Delta,t}
=
\mathcal D^{\otimes t}\qty(T_{\mu_{\mathcal H},t}).
\end{equation}
The map $\mathcal D^{\otimes t}$ is itself a pinching: it removes every matrix entry that is off-diagonal in the $t$-fold product basis. Hence Eq.~\eqref{eq:pinching_contraction} gives
\begin{align}
\norm{T_{\mu_\Delta,t}-T_{\nu_\Delta,t}}_p
&=
\norm{\mathcal D^{\otimes t}\qty(
T_{\mu_{\mathcal H},t}-T_{\nu_{\mathcal H},t})}_p\\
&\leq
\norm{T_{\mu_{\mathcal H},t}-T_{\nu_{\mathcal H},t}}_p.
\end{align}
This is Theorem~\ref{thm:moment_linear_pushforward} with an induced norm equal to one. Equality of the induced norm follows, for example, because $\mathcal D^{\otimes t}$ fixes every diagonal rank-one projector.
\end{proof}

The contraction constant in Theorem~\ref{thm:simplex_from_projective} is stronger than the generic channel-rank estimate. The latter would retain a factor depending on $d$, whereas dephasing is a conditional expectation and therefore cannot increase any Schatten norm.

\begin{rem}[Frame-potential estimate]
\label{rem:frame_potential_simplex}
Let $S_{\mathcal H}=\{\ket{\psi_i}\}_{i=1}^m$ be a uniform finite ensemble and set
\begin{equation}
F_t(S_{\mathcal H})
=
\frac{1}{m^2}\sum_{i,j=1}^m
|\braket{\psi_i}{\psi_j}|^{2t},
\qquad
D_t=\binom{d+t-1}{t}.
\end{equation}
The Welch bound is $F_t(S_{\mathcal H})\geq D_t^{-1}$, with equality exactly for a projective $t$-design \cite{DATTA20122455}. Write $T_\nu=T_{\nu_{\mathcal H},t}$ and $\Pi=\Pi_{\mathcal H_d,t}$. Since $T_\nu$ has trace one and is supported on the range of~$\Pi$,
\begin{equation}
\Tr(T_\nu\Pi)=\Tr T_\nu=1,
\qquad
\Tr\Pi=D_t.
\end{equation}
Moreover,
\begin{equation}
\label{eq:frame_potential_frobenius}
\norm{T_{\nu_{\mathcal H},t}-T_{\mu_{\mathcal H},t}}_2^2
=
F_t(S_{\mathcal H})-\frac{1}{D_t}.
\end{equation}
Indeed, $T_{\mu_{\mathcal H},t}=\Pi/D_t$ and $F_t(S_{\mathcal H})=\Tr(T_\nu^2)$, so direct expansion gives
\begin{align}
\norm{T_\nu-\Pi/D_t}_2^2
&=\Tr(T_\nu^2)-\frac{2}{D_t}\Tr(T_\nu\Pi)
+\frac{1}{D_t^2}\Tr\Pi\\
&=F_t(S_{\mathcal H})-\frac{1}{D_t}.
\end{align}

The difference in Eq.~\eqref{eq:frame_potential_frobenius} is Hermitian, traceless, and supported on a space of dimension $D_t$. If $Z$ has these properties, then
\begin{equation}
\norm{Z}_\infty
\leq
\sqrt{\frac{D_t-1}{D_t}}\norm{Z}_2.
\end{equation}
If $D_t=1$, then tracelessness forces $Z=0$. Suppose $D_t\geq2$, let $z_1,\ldots,z_{D_t}$ be the eigenvalues of $Z$, and choose $k$ such that $|z_k|=\norm{Z}_\infty$. Since $\sum_i z_i=0$, Cauchy--Schwarz gives
\begin{align}
|z_k|^2
&=\qty|\sum_{i\neq k}z_i|^2
\leq(D_t-1)\sum_{i\neq k}|z_i|^2\\
&=(D_t-1)\qty(\norm{Z}_2^2-|z_k|^2).
\end{align}
Rearranging proves the displayed spectral bound. Combining it with Theorem~\ref{thm:simplex_from_projective} gives
\begin{equation}
\norm{T_{\mu_\Delta,t}-T_{\nu_\Delta,t}}_\infty
\leq
W(S_{\mathcal H}),
\end{equation}
where
\begin{equation}
W(S_{\mathcal H})
:=
\sqrt{
\frac{D_t-1}{D_t}
\qty(F_t(S_{\mathcal H})-\frac{1}{D_t})}.
\end{equation}
Thus the simplex error can be estimated directly from the Gram matrix of the original vectors.
\end{rem}

\subsection{A direct bound for mixed-state designs}
\label{subsec:direct_mixed_state_designs}

Let $\mathcal H_{AB}=\mathcal H_A\otimes\mathcal H_B$. The partial trace sends a pure bipartite state to the reduced density operator
\begin{equation}
\ket\psi\bra\psi
\longmapsto
\rho_A=\Tr_B\ket\psi\bra\psi.
\end{equation}
It pushes the Fubini--Study measure on $\mathbb P(\mathcal H_A\otimes\mathcal H_B)$ forward to an induced measure on~$\mathcal B(\mathcal H_A)$ \cite{zyczkowski2001induced}. Exact mixed-state designs obtained in this way were studied in Ref.~\cite{czartowski2019isoentangled}. If $\nu_{\mathcal H}$ is another probability measure on the same projective space, denote the two pushforwards by
\begin{equation}
\mu_{\mathcal B}:=(\Tr_B)_*\mu_{\mathcal H},
\qquad
\nu_{\mathcal B}:=(\Tr_B)_*\nu_{\mathcal H},
\end{equation}
where the notation $\Tr_B$ on rays means $[\psi]\mapsto\Tr_B\ket\psi\bra\psi$.

\begin{obs}[Direct partial-trace bound]
\label{prop:naive_mixed_state_bound}
Let $\nu_{\mathcal H}$ be a $\delta_p$-approximate complex projective $t$-design on $\mathbb P(\mathcal H_A\otimes\mathcal H_B)$ relative to the Fubini--Study measure $\mu_{\mathcal H}$, and let $\nu_{\mathcal B}$ be the induced ensemble of reduced states. Then
\begin{equation}
\label{eq:direct_mixed_bound}
\delta'_p
=
\norm{T_{\mu_{\mathcal B},t}-T_{\nu_{\mathcal B},t}}_p
\leq
d_B^{t(1-1/p)}
\norm{T_{\mu_{\mathcal H},t}-T_{\nu_{\mathcal H},t}}_p
\leq
d_B^{t(1-1/p)}\delta_p.
\end{equation}
\end{obs}

\begin{proof}
The canonical order in the projective moment is $(\mathcal H_A\otimes\mathcal H_B)^{\otimes t}$. A fixed unitary permutation of tensor factors identifies it with $\mathcal H_A^{\otimes t}\otimes\mathcal H_B^{\otimes t}$; this regrouping preserves every Schatten norm. In the regrouped order, the pointwise identity
\begin{equation}
\qty(\Tr_B\ket\psi\bra\psi)^{\otimes t}
=
\Tr_{B_1\cdots B_t}\qty[
\qty(\ket\psi\bra\psi)^{\otimes t}]
\end{equation}
implies
\begin{equation}
T_{\nu_{\mathcal B},t}
=
\Tr_{B_1\cdots B_t}T_{\nu_{\mathcal H},t},
\qquad
T_{\mu_{\mathcal B},t}
=
\Tr_{B_1\cdots B_t}T_{\mu_{\mathcal H},t}.
\end{equation}
The traced subsystem has dimension $d_B^t$. Applying Fact~\ref{fact:partial_trace_bound} to the moment difference proves Eq.~\eqref{eq:direct_mixed_bound}.
\end{proof}

\subsection{A direct bound for channel designs}
\label{subsec:direct_channel}

A bipartite unitary does not specify a channel on subsystem $A$ until the initial state of $B$ is fixed. We use a maximally mixed environment because it leads to the $d_B^2$ environmental factor considered below. For $U\in\mathcal U(d_Ad_B)$, define
\begin{equation}
\label{eq:reduced_channel_definition}
\Phi_U(Q)
=
\Tr_B\qty[U\qty(Q\otimes\frac{\mathbb I_B}{d_B})U^\dagger].
\end{equation}
This channel is trace-preserving and unital, so the pushforward takes values in the set of bistochastic channels on $\mathcal H_A$.

Let $A'$ and $B'$ be reference copies of $A$ and $B$, and write
\begin{equation}
\ket{\Omega_A}=\frac{1}{\sqrt{d_A}}\sum_{i=1}^{d_A}\ket i_{A'}\ket i_A,
\qquad
\ket{\Omega_B}=\frac{1}{\sqrt{d_B}}\sum_{j=1}^{d_B}\ket j_{B'}\ket j_B.
\end{equation}
Under the canonical regrouping $A'\otimes A\otimes B'\otimes B\simeq A'\otimes B'\otimes A\otimes B$, set
\begin{equation}
\ket{\Omega_{AB}}
:=
\frac{1}{\sqrt{d_Ad_B}}
\sum_{i=1}^{d_A}\sum_{j=1}^{d_B}
\ket i_{A'}\ket j_{B'}\ket i_A\ket j_B.
\end{equation}
The normalized operator vector and its rank-one state are
\begin{equation}
\opket{U}
:=
\qty(\mathbb I_{A'B'}\otimes U)
\ket{\Omega_{AB}},
\qquad
\omega_U:=\opket{U}\opbra{U}.
\end{equation}
The normalized Choi state of $\Phi_U$ is
\begin{equation}
\label{eq:reduced_choi_state}
\omega_{\Phi_U}
=
(\id_{A'}\otimes\Phi_U)\qty(\ket{\Omega_A}\bra{\Omega_A})
=
\Tr_{B'B}\omega_U.
\end{equation}
Here is the normalization check in detail. In the regrouped tensor order,
\begin{equation}
\omega_U
=
(\mathbb I_{A'B'}\otimes U)
\ket{\Omega_{AB}}\bra{\Omega_{AB}}
(\mathbb I_{A'B'}\otimes U^\dagger).
\end{equation}
By definition, $\ket{\Omega_{AB}}$ is the image of $\ket{\Omega_A}\otimes\ket{\Omega_B}$ under the fixed regrouping from $A'AB'B$ to $A'B'AB$. Because $U$ acts trivially on the reference system $B'$ and
$\Tr_{B'}\ket{\Omega_B}\bra{\Omega_B}=\mathbb I_B/d_B$, tracing first over $B'$ and then over $B$ gives
\begin{align}
\Tr_{B'B}\omega_U
&=
\Tr_B\left[
(\mathbb I_{A'}\otimes U)
\qty(\ket{\Omega_A}\bra{\Omega_A}\otimes\frac{\mathbb I_B}{d_B})
(\mathbb I_{A'}\otimes U^\dagger)
\right]\\
&=(\id_{A'}\otimes\Phi_U)
\qty(\ket{\Omega_A}\bra{\Omega_A}),
\end{align}
which proves Eq.~\eqref{eq:reduced_choi_state}. Since $\Phi_U$ is trace-preserving and unital, respectively, its normalized Choi state obeys
\begin{equation}
\Tr_A\omega_{\Phi_U}=\frac{\mathbb I_{A'}}{d_A},
\qquad
\Tr_{A'}\omega_{\Phi_U}=\frac{\mathbb I_A}{d_A}.
\end{equation}

For a measure $\lambda$ on $\mathcal U(d_Ad_B)$, define its normalized Choi-vector moment by
\begin{equation}
\label{eq:choi_unitary_moment}
\widehat T_{\lambda,t}
=
\int\omega_U^{\otimes t}\,d\lambda(U).
\end{equation}
The corresponding channel moment is
\begin{equation}
T^{\mathcal C}_{\lambda,t}
=
\int\omega_{\Phi_U}^{\otimes t}\,d\lambda(U).
\end{equation}

\begin{prop}[Direct channel bound]
\label{prop:direct_channel_bound}
Let $\mu$ be Haar measure on $\mathcal U(d_Ad_B)$ and let $\nu$ be another probability measure. If
\begin{equation}
\widehat\delta_p
:=
\norm{\widehat T_{\mu,t}-\widehat T_{\nu,t}}_p,
\end{equation}
then the induced channel ensemble satisfies
\begin{equation}
\label{eq:direct_channel_bound}
\norm{T^{\mathcal C}_{\mu,t}-T^{\mathcal C}_{\nu,t}}_p
\leq
d_B^{2t(1-1/p)}\widehat\delta_p.
\end{equation}
\end{prop}

\begin{proof}
Equation~\eqref{eq:reduced_choi_state} gives
\begin{equation}
T^{\mathcal C}_{\mu,t}-T^{\mathcal C}_{\nu,t}
=
\Tr_{(B'B)^{\otimes t}}
\qty(\widehat T_{\mu,t}-\widehat T_{\nu,t}).
\end{equation}
Passing from $(A'B'AB)^{\otimes t}$ to $(A'A)^{\otimes t}\otimes(B'B)^{\otimes t}$ uses only a fixed permutation of tensor factors and hence does not change any Schatten norm. Each copy of $B'B$ has dimension $d_B^2$, so the entire traced subsystem has dimension $d_B^{2t}$. Fact~\ref{fact:partial_trace_bound} proves the claim.
\end{proof}

\begin{rem}[Conversion from the standard unitary moment]
\label{rem:realignment_conversion}
Let $D=d_Ad_B$ and let $T^{\mathrm U}_{\lambda,t}$ be the standard moment from Eq.~\eqref{eq:standard_unitary_moment}. To specify the realignment, write $U_{\alpha i}=\mel{\alpha}{U}{i}$ in a fixed basis of $\mathbb C^D$. The indices $i$ and $\alpha$ each range over the full $AB$ basis; in the channel decomposition they split as $i=(i_A,i_B)$ and $\alpha=(\alpha_A,\alpha_B)$. For one copy, define the entry permutation $\mathcal R_1$ by
\begin{equation}
\label{eq:realignment_components}
\qty[\mathcal R_1(Z)]_{(i,\alpha),(j,\beta)}
:=
Z_{(\alpha,\beta),(i,j)}.
\end{equation}
Since
\begin{equation}
\qty[U\otimes\overline U]_{(\alpha,\beta),(i,j)}
=U_{\alpha i}\overline{U_{\beta j}},
\end{equation}
whereas
\begin{equation}
\qty[\omega_U]_{(i,\alpha),(j,\beta)}
=\frac{1}{D}U_{\alpha i}\overline{U_{\beta j}},
\end{equation}
we have $\mathcal R_1(U\otimes\overline U)=D\omega_U$. For $t$ copies, first apply the fixed unitary permutations that place the factors in copywise order and then apply $\mathcal R_1$ to every copy. The resulting entry permutation $\mathcal R_t$ satisfies
\begin{equation}
\widehat T_{\lambda,t}
=
D^{-t}\mathcal R_t\qty(T^{\mathrm U}_{\lambda,t}).
\end{equation}
Realignment is an isometry for the Hilbert--Schmidt norm, but not for general Schatten norms. For example, under the one-copy realignment, the identity on a $D^2$-dimensional tensor-product space can be sent to the rank-one outer product of the unnormalized vectorization of $\mathbb I_D$. Its spectral norm is $D$, whereas that of the original identity is $1$.

Both sides of $\mathcal R_t$ are matrices of size $N=D^{2t}$. If $1\leq p\leq2$, the Schatten-norm comparisons and Hilbert--Schmidt invariance give
\begin{align}
\norm{\mathcal R_t(Z)}_p
&\leq N^{1/p-1/2}\norm{\mathcal R_t(Z)}_2\\
&=N^{1/p-1/2}\norm{Z}_2
\leq N^{1/p-1/2}\norm{Z}_p.
\end{align}
If $2\leq p\leq\infty$, they instead give
\begin{align}
\norm{\mathcal R_t(Z)}_p
&\leq\norm{\mathcal R_t(Z)}_2
=\norm{Z}_2\\
&\leq N^{1/2-1/p}\norm{Z}_p.
\end{align}
Combining the two cases yields
\begin{equation}
\norm{\mathcal R_t(Z)}_p
\leq
N^{|1/p-1/2|}\norm{Z}_p
=
D^{t|1-2/p|}\norm{Z}_p.
\end{equation}
Therefore, if
\begin{equation}
\delta_p^{\mathrm U}
:=
\norm{T^{\mathrm U}_{\mu,t}-T^{\mathrm U}_{\nu,t}}_p,
\end{equation}
then
\begin{equation}
\label{eq:realignment_error_conversion}
\widehat\delta_p
\leq
D^{t(|1-2/p|-1)}\delta_p^{\mathrm U}.
\end{equation}
Exact unitary-design conditions are equivalent in the two conventions because realignment is invertible. Approximation errors, however, cannot be silently identified for any $p$. At $p=2$, entry realignment is isometric and the only conversion is the explicit normalization factor $D^{-t}$; for general $p$, Eq.~\eqref{eq:realignment_error_conversion} gives the required comparison.
\end{rem}

\section{Structural refinements}
\label{sec:refined_applications}

The estimates in Section~\ref{sec:direct_applications} use the norm of a map on its full ambient operator space. Projective and Choi-vector moments occupy a much smaller subspace: they are supported on the totally symmetric subspace across the $t$ copies. The resulting restriction does not simply make the environmental tensor factors symmetric. Instead, it changes the induced norm of the partial trace. The following proposition identifies that norm exactly.

\subsection{Partial trace on symmetric input support}

\begin{prop}[Sharp symmetric-support partial-trace bound]
\label{thm:symmetric_ptrace}
Let $\mathcal K$ and $\mathcal E$ be finite-dimensional Hilbert spaces, let $d_{\mathcal{E}}=\dim\mathcal E$, and let
\begin{equation}
S=\Sym_t(\mathcal K\otimes\mathcal E).
\end{equation}
After regrouping the factors as $\mathcal K^{\otimes t}\otimes\mathcal E^{\otimes t}$, let $\Pi_S$ denote the projector onto $S$. If $X=\Pi_SX\Pi_S$, then, for every $1\leq p\leq\infty$,
\begin{equation}
\label{eq:symmetric_ptrace_bound}
\norm{\Tr_{\mathcal E^{\otimes t}}X}_p
\leq
D_{\mathcal E,t}^{1-1/p}\norm{X}_p,
\qquad
D_{\mathcal E,t}=\binom{d_{\mathcal{E}}+t-1}{t}.
\end{equation}
The constant is sharp.
\end{prop}

\begin{proof}
Identify $\mathcal L(S)$ with the corner
\begin{equation}
\Pi_S\mathcal L(\mathcal K^{\otimes t}\otimes\mathcal E^{\otimes t})\Pi_S.
\end{equation}
The partial trace restricted to this corner is the map
\begin{equation}
\Lambda:\mathcal L(S)\to\mathcal L(\mathcal K^{\otimes t}),
\qquad
\Lambda(X)=\Tr_{\mathcal E^{\otimes t}}X.
\end{equation}
This map is completely positive and trace-preserving. Its Hilbert--Schmidt adjoint, that is the adjoint with respect to Hilbert--Schmidt inner product, is
\begin{equation}
\Lambda^*(Y)=\Pi_S\qty(Y\otimes\mathbb I_{\mathcal E^{\otimes t}})\Pi_S.
\end{equation}
In particular, $\Lambda^*(\mathbb I_{\mathcal K^{\otimes t}})=\Pi_S$, the identity of the corner algebra. Thus $\Lambda^*$ is unital and positive, and the positive-map norm identity (equivalently, the Russo--Dye theorem) gives $\norm{\Lambda^*}_{\infty\to\infty}=1$. Duality between induced norms of a map and its adjoint now yields
\begin{equation}
\norm{\Lambda}_{1\to1}=1.
\end{equation}
At the other endpoint, $\Lambda$ is positive. Applying the same positive-map norm identity to $\Lambda$ gives
\begin{equation}
\norm{\Lambda}_{\infty\to\infty}
=
\norm{\Lambda(\Pi_S)}_\infty.
\end{equation}

The symmetric projector~\eqref{eq:symmetric_projector}, after regrouping the $\mathcal K$ and $\mathcal E$ factors, is
\begin{equation}
\Pi_S
=
\frac{1}{t!}\sum_{\sigma\in S_t}
P_{\mathcal K}(\sigma)\otimes P_{\mathcal E}(\sigma).
\end{equation}
If $c(\sigma)$ denotes the number of cycles of $\sigma$, then
\begin{equation}
\Tr P_{\mathcal E}(\sigma)=d_{\mathcal{E}}^{c(\sigma)}.
\end{equation}
Indeed, in a product basis a diagonal matrix element of $P_{\mathcal E}(\sigma)$ is nonzero precisely when the basis label is constant along each cycle of $\sigma$; each cycle admits $d_{\mathcal{E}}$ independent labels.
Consequently,
\begin{equation}
\label{eq:partial_trace_symmetric_projector}
\Lambda(\Pi_S)
=
\frac{1}{t!}\sum_{\sigma\in S_t}
d_{\mathcal{E}}^{c(\sigma)}P_{\mathcal K}(\sigma).
\end{equation}
Every permutation operator has spectral norm one. The triangle inequality and the cycle-index identity therefore give
\begin{align}
\norm{\Lambda(\Pi_S)}_\infty
&\leq
\frac{1}{t!}\sum_{\sigma\in S_t}d_{\mathcal{E}}^{c(\sigma)}\\
&=
\frac{d_{\mathcal{E}}(d_{\mathcal{E}}+1)\cdots(d_{\mathcal{E}}+t-1)}{t!}\\
&=
\binom{d_{\mathcal{E}}+t-1}{t}
=D_{\mathcal E,t}.
\end{align}
For completeness, the cycle-index identity follows from a short recurrence. Set
\begin{equation}
a_t(e):=\sum_{\sigma\in S_t}d_{\mathcal{E}}^{c(\sigma)}.
\end{equation}
When the symbol $t$ is adjoined to a permutation of $t-1$ symbols, it either forms a new one-element cycle, contributing $d_{\mathcal{E}}\,a_{t-1}(d_{\mathcal{E}})$, or it is inserted immediately after one of the $t-1$ existing symbols, contributing $(t-1)a_{t-1}(d_{\mathcal{E}})$. Hence
\begin{equation}
a_t(d_{\mathcal{E}})=(d_{\mathcal{E}}+t-1)a_{t-1}(d_{\mathcal{E}}),
\qquad
a_0(d_{\mathcal{E}})=1.
\end{equation}
It follows that $a_t(d_{\mathcal{E}})=d_{\mathcal{E}}(d_{\mathcal{E}}+1)\cdots(d_{\mathcal{E}}+t-1)$, as used above.
This upper bound is attained. Indeed, for every unit vector $\ket v\in\mathcal K$, the vector $\ket v^{\otimes t}$ is fixed by every $P_{\mathcal K}(\sigma)$ and is therefore an eigenvector of Eq.~\eqref{eq:partial_trace_symmetric_projector} with eigenvalue $D_{\mathcal E,t}$. Hence
\begin{equation}
\norm{\Lambda}_{\infty\to\infty}=D_{\mathcal E,t}.
\end{equation}

Complex interpolation between the Schatten classes $S_1$ and $S_\infty$ now yields
\begin{equation}
\norm{\Lambda}_{p\to p}
\leq
\norm{\Lambda}_{1\to1}^{1/p}
\norm{\Lambda}_{\infty\to\infty}^{1-1/p}
=
D_{\mathcal E,t}^{1-1/p},
\end{equation}
which proves Eq.~\eqref{eq:symmetric_ptrace_bound}.

To see sharpness for every $p$, note that
\begin{equation}
\Sym_t(\mathcal K)\otimes\Sym_t(\mathcal E)
\subseteq
\Sym_t(\mathcal K\otimes\mathcal E)
\end{equation}
under the same regrouping. Choose any nonzero operator $Y$ on $\Sym_t(\mathcal K)$ and set
\begin{equation}
X=Y\otimes\mathbb I_{\Sym_t(\mathcal E)}.
\end{equation}
More precisely, extend $Y$ by zero from $\Sym_t(\mathcal K)$ to $\mathcal K^{\otimes t}$. The displayed formula then defines $X$ on $\Sym_t(\mathcal K)\otimes\Sym_t(\mathcal E)$, with $X$ equal to zero on its orthogonal complement inside $S$. As an operator on the full regrouped tensor product, $X$ is the tensor product of the extended $Y$ with the orthogonal projector onto $\Sym_t(\mathcal E)$. Therefore
\begin{equation}
\norm{X}_p
=
D_{\mathcal E,t}^{1/p}\norm{Y}_p,
\qquad
\norm{\Lambda(X)}_p
=
D_{\mathcal E,t}\norm{Y}_p,
\end{equation}
so the ratio is exactly $D_{\mathcal E,t}^{1-1/p}$.
\end{proof}

\begin{rem}[Why the proof is not a Kraus-rank reduction]
The global symmetry of $S=\Sym_t(\mathcal K\otimes\mathcal E)$ does not imply that the $\mathcal E$ factors lie in $\Sym_t(\mathcal E)$. For example,
\begin{equation}
\Sym_2(\mathcal K\otimes\mathcal E)
\simeq
\qty(\Sym_2(\mathcal K)\otimes\Sym_2(\mathcal E))
\oplus
\qty(\wedge^2\mathcal K\otimes\wedge^2\mathcal E).
\end{equation}
Antisymmetric environmental modes therefore occur. For $t=2$, Eq.~\eqref{eq:partial_trace_symmetric_projector} becomes
\begin{equation}
\Tr_{\mathcal E^{\otimes2}}\Pi_S
=
\binom{d_{\mathcal{E}}+1}{2}\Pi_{\Sym_2(\mathcal K)}
+
\binom{d_{\mathcal{E}}}{2}\Pi_{\wedge^2\mathcal K}.
\end{equation}
To derive this formula directly, let $F_{\mathcal K}$ and $F_{\mathcal E}$ be the two-copy swap operators. Then
\begin{equation}
\Pi_S=\frac{1}{2}\qty(\mathbb I+F_{\mathcal K}\otimes F_{\mathcal E}),
\end{equation}
Since both swaps have eigenvalues $\pm1$, the $+1$ eigenspace of $F_{\mathcal K}\otimes F_{\mathcal E}$ consists exactly of the $(+,+)$ and $(-,-)$ sectors, which gives the decomposition above. Moreover,
\begin{align}
\Tr_{\mathcal E^{\otimes2}}\Pi_S
&=\frac{1}{2}\qty(d_{\mathcal{E}}^2\mathbb I_{\mathcal K^{\otimes2}}+d_{\mathcal{E}}F_{\mathcal K})\\
&=\frac{d_{\mathcal{E}}(d_{\mathcal{E}}+1)}{2}\Pi_{\Sym_2(\mathcal K)}
+\frac{d_{\mathcal{E}}(d_{\mathcal{E}}-1)}{2}\Pi_{\wedge^2\mathcal K}.
\end{align}
The largest eigenvalue is nevertheless $\binom{d_{\mathcal{E}}+1}{2}$. Thus Proposition~\ref{thm:symmetric_ptrace} is a restricted induced-norm statement, not a claim that the restricted channel has only $D_{\mathcal E,t}$ Kraus operators.
\end{rem}

\subsection{Refined bounds for mixed-state designs}
\label{subsec:refined_mixed_state_designs}

The projective moment difference
\begin{equation}
X
=
T_{\mu_{\mathcal H},t}-T_{\nu_{\mathcal H},t}
\end{equation}
is supported on
\begin{equation}
\Sym_t(\mathcal H_A\otimes\mathcal H_B).
\end{equation}
Indeed, every integrand has the form
\begin{equation}
\qty(\ket\psi\bra\psi)^{\otimes t}
=
\ket{\psi^{\otimes t}}\bra{\psi^{\otimes t}},
\end{equation}
and $\ket{\psi^{\otimes t}}$ is fixed by every permutation of the $t$ copies. Both moments, and hence their difference, vanish on the orthogonal complement of the symmetric subspace.
This is the additional information ignored by Observation~\ref{prop:naive_mixed_state_bound}.

\begin{thm}[Symmetry-refined mixed-state bound]
\label{thm:improved_bound_states}
With the notation of Observation~\ref{prop:naive_mixed_state_bound}, set
\begin{equation}
D_{B,t}=\binom{d_B+t-1}{t}.
\end{equation}
Then, for every $1\leq p\leq\infty$,
\begin{equation}
\label{eq:refined_mixed_bound}
\delta'_p
\leq
D_{B,t}^{1-1/p}
\norm{T_{\mu_{\mathcal H},t}-T_{\nu_{\mathcal H},t}}_p
\leq
D_{B,t}^{1-1/p}\delta_p.
\end{equation}
\end{thm}

\begin{proof}
As in the proof of Observation~\ref{prop:naive_mixed_state_bound},
\begin{equation}
T_{\mu_{\mathcal B},t}-T_{\nu_{\mathcal B},t}
=
\Tr_{B_1\cdots B_t}
\qty(T_{\mu_{\mathcal H},t}-T_{\nu_{\mathcal H},t}).
\end{equation}
The operator inside the partial trace has support on $\Sym_t(\mathcal H_A\otimes\mathcal H_B)$. Apply Proposition~\ref{thm:symmetric_ptrace} with $\mathcal K=\mathcal H_A$ and $\mathcal E=\mathcal H_B$.
\end{proof}

For fixed $t$ and $d_B\to\infty$,
\begin{equation}
D_{B,t}
=
\binom{d_B+t-1}{t}
=
\frac{d_B^t}{t!}+O(d_B^{t-1}).
\end{equation}
Since $\Sym_t(\mathcal H_B)$ is a subspace of $\mathcal H_B^{\otimes t}$, one has $D_{B,t}\leq d_B^t$ for every $d_B$ and $t$. The refined coefficient is therefore never larger than the direct coefficient. Their exact ratio, direct divided by refined, is
\begin{equation}
\qty(\frac{d_B^t}{D_{B,t}})^{1-1/p},
\end{equation}
which tends to $(t!)^{1-1/p}$ as $d_B\to\infty$ with $t$ fixed.

The same argument gives a transparent family of mixed-norm estimates.

\begin{prop}[Mixed-norm consequence]
\label{prop:mixed_norm_bound_states}
Let $\nu_{\mathcal H}$ be a $\delta_p$-approximate complex projective $t$-design relative to $\mu_{\mathcal H}$, and let $\mu_{\mathcal B}$ and $\nu_{\mathcal B}$ be their partial-trace pushforwards. Set
\begin{equation}
D_{AB,t}=\binom{d_Ad_B+t-1}{t}.
\end{equation}
For $1\leq p,q\leq\infty$, define
\begin{equation}
\delta'_q
:=
\norm{T_{\mu_{\mathcal B},t}-T_{\nu_{\mathcal B},t}}_q.
\end{equation}
Then
\begin{equation}
\label{eq:mixed_norm_consequence}
\delta'_q
\leq
D_{B,t}^{1-1/q}
D_{AB,t}^{\max\{0,1/q-1/p\}}\delta_p.
\end{equation}
\end{prop}

\begin{proof}
The input difference $X$ is supported on a space of dimension $D_{AB,t}$, so $\operatorname{rank}(X)\leq D_{AB,t}$. First apply Proposition~\ref{thm:symmetric_ptrace} in Schatten $q$-norm and then Eq.~\eqref{eq:schatten_comparison}:
\begin{align}
\norm{\Tr_{B_1\cdots B_t}X}_q
&\leq
D_{B,t}^{1-1/q}\norm{X}_q\\
&\leq
D_{B,t}^{1-1/q}
D_{AB,t}^{\max\{0,1/q-1/p\}}\norm{X}_p.
\end{align}
Using $\norm{X}_p\leq\delta_p$ proves the result.
\end{proof}

Taking $q=p$ recovers Theorem~\ref{thm:improved_bound_states}. If $p\leq q$, then $1/q-1/p\leq0$, so the factor involving $D_{AB,t}$ disappears. Proposition~\ref{prop:mixed_norm_bound_states} is a convenient consequence of two general inequalities; no claim of sharpness is made when $p\neq q$.

\subsection{Refined bounds for channel designs}
\label{subsec:refined_channel_designs}

After the fixed regrouping used in the proof of Proposition~\ref{prop:direct_channel_bound}, the normalized Choi-vector moment~\eqref{eq:choi_unitary_moment} is a projective moment on the Hilbert space
\begin{equation}
\qty(\mathcal H_{A'}\otimes\mathcal H_A)
\otimes
\qty(\mathcal H_{B'}\otimes\mathcal H_B).
\end{equation}
Its difference is therefore supported on the totally symmetric subspace across the $t$ copies. The first factor has dimension $d_A^2$ and the environmental factor $B'B$ has dimension $d_B^2$.
Explicitly, every integrand is
\begin{equation}
\omega_U^{\otimes t}
=
\qty(\opket{U})^{\otimes t}\qty(\opbra{U})^{\otimes t},
\end{equation}
and the vector $\opket{U}^{\otimes t}$ is invariant under simultaneous permutations of its $t$ copies.

\begin{thm}[Symmetry-refined channel bound]
\label{thm:channel_design_bound}
Let $\mu$ be Haar measure on $\mathcal U(d_Ad_B)$, let $\nu$ be another probability measure, and define $\widehat\delta_p$ using the normalized Choi-vector moments as in Proposition~\ref{prop:direct_channel_bound}. Then the induced ensemble of bistochastic channels~\eqref{eq:reduced_channel_definition} satisfies
\begin{equation}
\label{eq:refined_channel_bound}
\norm{T^{\mathcal C}_{\mu,t}-T^{\mathcal C}_{\nu,t}}_p
\leq
\binom{d_B^2+t-1}{t}^{1-1/p}\widehat\delta_p.
\end{equation}
In terms of the standard unitary-moment error $\delta_p^{\mathrm U}$,
\begin{equation}
\label{eq:refined_channel_standard_convention}
\norm{T^{\mathcal C}_{\mu,t}-T^{\mathcal C}_{\nu,t}}_p
\leq
(d_Ad_B)^{-t+t|1-2/p|}
\binom{d_B^2+t-1}{t}^{1-1/p}
\delta_p^{\mathrm U}.
\end{equation}
\end{thm}

\begin{proof}
Equation~\eqref{eq:reduced_choi_state} shows that the channel moment difference is obtained by tracing out $(B'B)^{\otimes t}$ from the normalized Choi-vector moment difference. The latter is supported on
\begin{equation}
\Sym_t\qty(
(\mathcal H_{A'}\otimes\mathcal H_A)
\otimes
(\mathcal H_{B'}\otimes\mathcal H_B)).
\end{equation}
Apply Proposition~\ref{thm:symmetric_ptrace} with an environment of dimension $d_B^2$ to obtain Eq.~\eqref{eq:refined_channel_bound}. Equation~\eqref{eq:refined_channel_standard_convention} then follows from the realignment estimate~\eqref{eq:realignment_error_conversion}.
\end{proof}

For fixed $t$ and large $d_B$, the refined channel dimension satisfies
\begin{equation}
\binom{d_B^2+t-1}{t}
=
\frac{d_B^{2t}}{t!}+O(d_B^{2t-2}),
\end{equation}
so, for $p>1$, the same asymptotic factor $(t!)^{1-1/p}$ separates the refined coefficient from the direct one in Eq.~\eqref{eq:direct_channel_bound}. At $p=1$, both coefficients equal one.
As in the mixed-state case, the symmetric dimension is never larger than the full environmental dimension, so the refined estimate is never worse. At $t=1$ the two dimensions agree, as they must.

\section{Numerical validation}
\label{sec:num}

\begin{figure}[H]
\centering
\includegraphics[width=\linewidth]{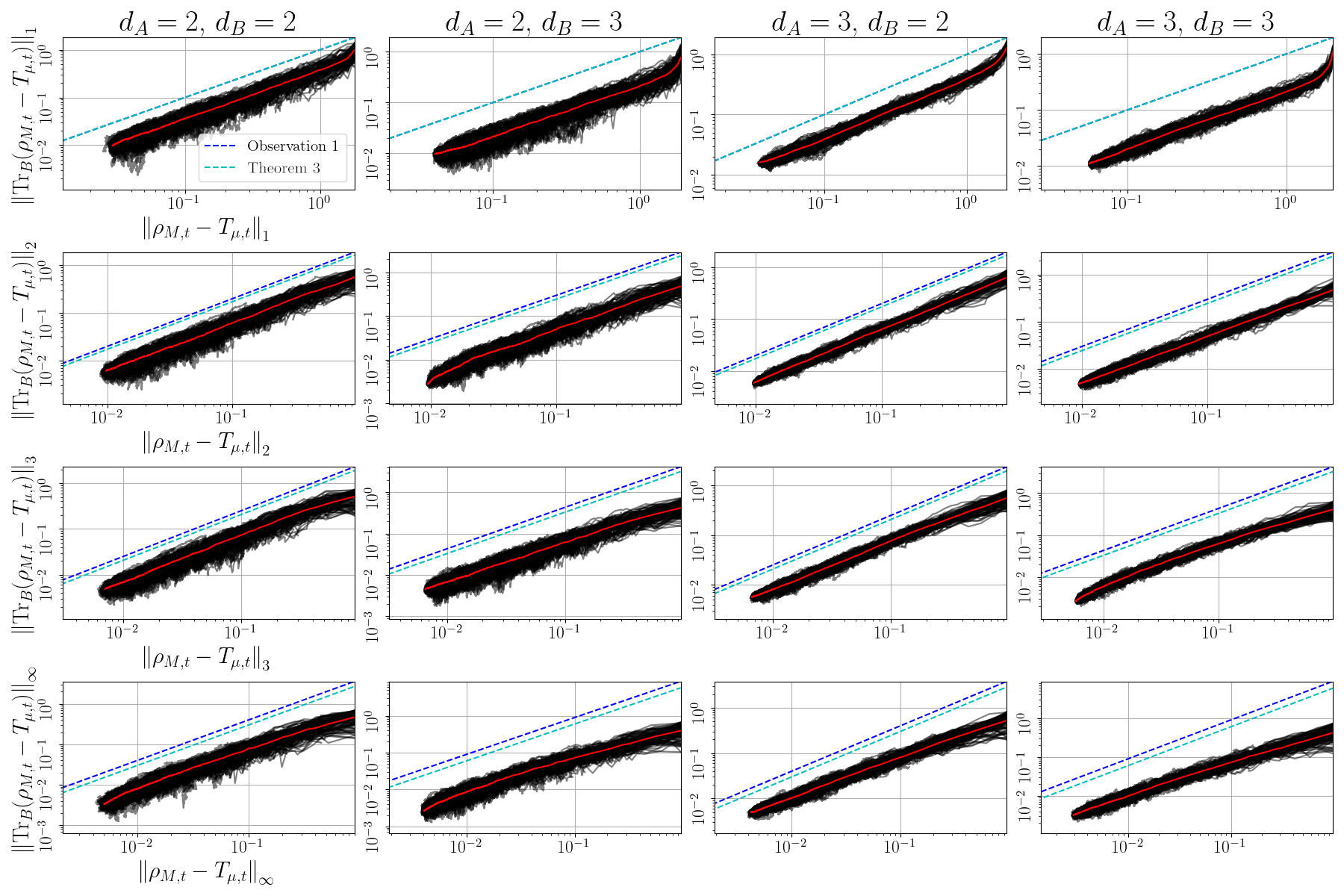}
\caption{Monte Carlo convergence of approximate mixed-state designs obtained by partially tracing empirical complex projective designs. The plots show the Schatten-$p$ errors in Eq.~\eqref{eq:numerical_error_pair} for $t=2$ and subsystem dimensions $d_A,d_B\in\{2,3\}$. The figure uses the shorthand $\rho_{M,t}=T^{(M)}_{\mathcal H,t}$ for the empirical projective moment and $\Tr_B$ for the tensor-power trace $\Tr_{B_1B_2}$. Each black curve is one trajectory with $1\leq M\leq10^4$, and the solid red curve is the average over $100$ repetitions. The dashed blue curves show the direct partial-trace estimate from Observation~\ref{prop:naive_mixed_state_bound}; the dashed cyan curves show the symmetry-refined estimate from Theorem~\ref{thm:improved_bound_states}.}
\label{fig:convergence_states}
\end{figure}

We numerically compare the direct mixed-state estimate from Observation~\ref{prop:naive_mixed_state_bound} with the symmetry-refined estimate from Theorem~\ref{thm:improved_bound_states}. The purpose of this experiment is not to verify the theorems, which hold deterministically, but to examine how closely the corresponding coefficients describe typical empirical moment errors.

For fixed subsystem dimensions $d_A$ and $d_B$, each repetition used a fresh sample of $N=10^4$ independent Haar-random pure states
\begin{equation}
\qty{\ket{\psi_i}\in\mathcal H_A\otimes\mathcal H_B}_{i=1}^{N}.
\end{equation}
Concretely, a Haar-random pure state was sampled by drawing a vector with independent standard complex Gaussian entries and normalizing it to unit length. This procedure is rotationally invariant and therefore induces the Fubini--Study distribution on projective space.
For each $1\leq M\leq N$, the empirical projective moment was
\begin{equation}
T^{(M)}_{\mathcal H,t}
=
\frac{1}{M}\sum_{j=1}^{M}
\qty(\ket{\psi_j}\bra{\psi_j})^{\otimes t}.
\end{equation}
The exact Fubini--Study moment is
\begin{equation}
T_{\mu_{\mathcal H},t}
=
\frac{\Pi_{\mathcal H_A\otimes\mathcal H_B,t}}
{\binom{d_Ad_B+t-1}{t}}.
\end{equation}

In parallel, we applied the tensor-power partial trace to obtain the empirical mixed-state moment. The corresponding exact induced moment is
\begin{equation}
T_{\mu_{\mathcal B},t}
=
\Tr_{B_1\cdots B_t}T_{\mu_{\mathcal H},t}.
\end{equation}
Thus, for each $M$, we evaluated
\begin{equation}
\label{eq:numerical_error_pair}
\qty(
\norm{T^{(M)}_{\mathcal H,t}-T_{\mu_{\mathcal H},t}}_p,
\norm{
\Tr_{B_1\cdots B_t}
\qty(T^{(M)}_{\mathcal H,t}-T_{\mu_{\mathcal H},t})}_p
),
\qquad
p\in\{1,2,3,\infty\}.
\end{equation}

The experiment was performed for all subsystem dimensions $d_A,d_B\in\{2,3\}$, with $100$ independent repetitions for each pair $(d_A,d_B)$. All Schatten norms were computed from the singular values of the corresponding moment differences. The results for $t=2$ are shown in Fig.~\ref{fig:convergence_states} on a log--log scale. Each black curve corresponds to one Monte Carlo trajectory over $1\leq M\leq10^4$, while the red curve shows the average over the $100$ repetitions.

The numerical results display the hierarchy predicted by the analytical estimates. In all tested cases, the coefficient from Theorem~\ref{thm:improved_bound_states} is no larger than the direct coefficient from Observation~\ref{prop:naive_mixed_state_bound}: it is strictly smaller for $p>1$, while the two coefficients agree for the trace norm, $p=1$, because both equal one. For $d_A=d_B=2$, the refined estimate is the closest of the displayed estimates to the observed upper envelope. For the other tested dimension pairs, the data leave room for further improvements tailored to the special form of moment differences, even though Proposition~\ref{thm:symmetric_ptrace} is sharp on the full symmetric-supported operator space.

\section{Concluding remarks}
\label{sec:conclusions}

We have developed a moment-operator framework for approximate pushforward designs. The main transfer theorem isolates the precise property required of a pushforward: the point map must induce a linear map between the chosen moment kernels. Once that map is identified, the output approximation error is controlled by its induced Schatten norm. This formulation covers ordinary linear maps on vectors, linear maps on operator-valued points, and maps such as the reduction of a bipartite unitary that become linear only after an appropriate Choi representation is chosen.

The applications naturally separate into two levels. The direct results use standard norm properties. Complete dephasing is a pinching and is contractive in every Schatten norm, so the approximation error of a simplex design is no larger than that of the projective design from which it is obtained. Ordinary partial-trace inequalities give immediate bounds for mixed-state designs and, after fixing a maximally mixed environment and normalized Choi convention, for ensembles of bistochastic channels.

The refined results use symmetric input support. We proved that partial trace on operators supported on $\Sym_t(\mathcal K\otimes\mathcal E)$ has sharp induced Schatten-$p$ norm
\begin{equation}
\binom{\dim\mathcal E+t-1}{t}^{1-1/p}.
\end{equation}
The proof is based on the partial trace of the symmetric projector and interpolation between the trace- and spectral-norm endpoints. Importantly, the improvement is not a reduction of the Kraus rank to the dimension of $\Sym_t(\mathcal E)$; globally symmetric tensors can contain nonsymmetric environmental sectors. Applying the correct restricted norm yields the improved mixed-state coefficient $\binom{d_B+t-1}{t}^{1-1/p}$ and, in the normalized Choi-vector convention, the channel coefficient $\binom{d_B^2+t-1}{t}^{1-1/p}$.

The channel application also illustrates why moment conventions must be stated explicitly for approximate designs. The standard unitary moment and the rank-one Choi-vector moment are related by an invertible entry realignment followed by the normalization factor $D^{-t}$. Entry realignment itself is a Hilbert--Schmidt isometry, while the normalized errors satisfy $\widehat\delta_2=D^{-t}\delta_2^{\mathrm U}$. For general $p$, Eq.~\eqref{eq:realignment_error_conversion} gives the corresponding comparison factor, which is trivial at the endpoints $p=1$ and $p=\infty$. The distinction is immaterial in the exact theory but essential for quantitative approximate bounds.

The numerical experiments for reduced random pure states are consistent with the analytical hierarchy. They also suggest a distinction between sharpness on the full symmetric-supported operator space and sharpness on the smaller set of differences that can arise from probability measures. Characterizing this latter set may lead to stronger constants for particular design families.

Several extensions remain open. One direction is to determine restricted induced norms for moment operators with symmetries other than total permutation symmetry. Another is to treat nonlinear point maps whose moment kernels do not admit a finite-order linear intertwiner. It would also be useful to compare systematically the approximation parameters induced by the standard unitary and normalized Choi representations.

\acknowledgments

\paragraph{Funding}
JCz acknowledges support by the start-up grant of the Nanyang Assistant Professorship at Nanyang Technological University in Singapore, awarded to Nelly Ng, and by Taighde {\'E}ireann -- Research Ireland under Grant No.~IRCLA/2022/3922. AS acknowledges support by the National Science Centre, Poland, under OPUS Grant No.~\nolinkurl{UMO2020/37/B/ST2/02478}. K{\.Z} acknowledges financial support from the European Union under ERC Advanced Grant \emph{TAtypic}, Project No.~101142236.

\paragraph{Data Availability Statement}
No external datasets were used. No fixed random seed was retained, so the individual Monte Carlo trajectories are not bit-for-bit reproducible. Their statistical behavior can be reproduced from the sampling procedure described in Section~\ref{sec:num}.

\paragraph{Declarations}

\paragraph{Conflicting Interests}
The authors declare no conflicts of interest relevant to the content of this article.

\bibliographystyle{quantum_abbr}
\bibliography{references-REVISED}

@article{scott2006tight,
  title={Tight informationally complete quantum measurements},
  author={Scott, Andrew J},
  journal={J. Phys. A: Math. Gen.},
  doi = {10.1088/0305-4470/39/43/009},
  volume={39},
  number={43},
  pages={13507},
  year={2006},
  publisher={IOP Publishing}
}

@article{scott2008optimizing,
  title={Optimizing quantum process tomography with unitary 2-designs},
  author={Scott, Andrew J},
  journal={J. Phys. A: Math. Theor.},
  doi = {10.1088/1751-8113/41/5/055308},
  volume={41},
  number={5},
  pages={055308},
  year={2008},
  publisher={IOP Publishing}
}

@article{dankert2009exact,
  title={Exact and approximate unitary 2-designs and their application to fidelity estimation},
  author={Dankert, Christoph and Cleve, Richard and Emerson, Joseph and Livine, Etera},
  journal={Phys. Rev. A},
  doi = {10.1103/PhysRevA.80.012304},
  volume={80},
  number={1},
  pages={012304},
  year={2009},
  publisher={APS}
}

@article{hoggar1982,
title = {$t$-Designs in Projective Spaces},
journal = {European Journal of Combinatorics},
volume = {3},
number = {3},
pages = {233-254},
year = {1982},
issn = {0195-6698},
doi = {https://doi.org/10.1016/S0195-6698(82)80035-8},
url = {https://www.sciencedirect.com/science/article/pii/S0195669882800358},
author = {S.G. Hoggar}
}

@INPROCEEDINGS{ambainis07,
  author={Ambainis, Andris and Emerson, Joseph},
  booktitle={Twenty-Second Annual IEEE Conference on Computational Complexity (CCC'07)}, 
  title={Quantum t-designs: t-wise Independence in the Quantum World}, 
  year={2007},
  volume={},
  number={},
  pages={129-140},
  doi={10.1109/CCC.2007.26}
  }

@article{iosue2023projective,
  author    = {Joseph T. Iosue and T. C. Mooney and Adam Ehrenberg and Alexey V. Gorshkov},
  title     = {Projective toric designs, quantum state designs, and mutually unbiased bases},
  journal   = {Quantum},
  volume    = {8},
  pages     = {1546},
  year      = {2024},
  url       = {https://doi.org/10.22331/q-2024-12-03-1546},
  doi       = {10.22331/Q-2024-12-03-1546}
}

@article{czartowski2019isoentangled,
  title     = {Iso-entangled {M}utually {U}nbiased {B}ases, symmetric quantum measurements and mixed-state designs},
  author    = {Jakub Czartowski and D. Goyeneche and M. Grassl and K. {\.Z}yczkowski},
  journal   = {Phys. Rev. Lett.},
  doi = {10.1103/PhysRevLett.124.090503},
  year      = {2020},
  volume={124},
  pages={090503},
  bibSource = {Semantic Scholar https://www.semanticscholar.org/paper/50b32face170554d268af7ace3567c7857d52415},
  url       = {https://arxiv.org/abs/1906.12291v2}
}

@article{gauss1815methodus,
  title={Methodus nova integralium valores per approximationem inveniendi},
  author={Gauss, Carl Friedrich},
  journal={Comm. Soc. Sci. G{\"o}ttingen Math.},
  volume={3},
  pages={29--76},
  year={1815},
  doi = {10.1017/CBO9781139058247.008},
}

@article{SEYMOUR1984213,
title = {Averaging sets: A generalization of mean values and spherical designs},
journal = {Adv. Math.},
doi = {10.1016/0001-8708(84)90022-7},
volume = {52},
number = {3},
pages = {213-240},
year = {1984},
issn = {0001-8708},
url = {https://www.sciencedirect.com/science/article/pii/0001870884900227},
author = {P.D Seymour and Thomas Zaslavsky}
}

@article{rastegin2012relations,
  title     = {Relations for certain symmetric norms and anti-norms before and after partial trace},
  author    = {Rastegin, Alexey E},
  journal   = {Journal of Statistical Physics},
  volume    = {148},
  pages     = {1040-1053},
  year      = {2012},
  publisher = {Springer},
  doi = {10.1007/s10955-012-0569-8}
}

@article{Delsarte1977,
  title = {Spherical codes and designs},
  volume = {6},
  ISSN = {1572-9168},
  url = {http://dx.doi.org/10.1007/BF03187604},
  DOI = {10.1007/bf03187604},
  number = {3},
  journal = {Geometriae Dedicata},
  publisher = {Springer Science and Business Media LLC},
  author = {Delsarte,  P. and Goethals,  J. M. and Seidel,  J. J.},
  year = {1977},
  month = sep,
  pages = {363--388}
}

@techreport{Neumaier1981,
  author = {A. Neumaier},
  title = {Combinatorial configurations in terms of distances},
  institution = {Eindhoven University of Technology},
  year = {1981},
  type = {Lecture Notes},
  number = {81-09}
}

@article{sawicki2017criteria,
  title = {Criteria for universality of quantum gates},
  author = {Sawicki, Adam and Karnas, Katarzyna},
  journal = {Phys. Rev. A},
  volume = {95},
  issue = {6},
  pages = {062303},
  numpages = {6},
  year = {2017},
  month = {Jun},
  publisher = {American Physical Society},
  doi = {10.1103/PhysRevA.95.062303},
  url = {https://link.aps.org/doi/10.1103/PhysRevA.95.062303}
}

@article{brandao2016local,
  title     = {Local random quantum circuits are approximate polynomial-designs},
  author    = {Brandao, Fernando GSL and Harrow, Aram W and Horodecki, Micha{\l}},
  journal   = {Communications in Mathematical Physics},
  volume    = {346},
  pages     = {397-434},
  year      = {2016},
  publisher = {Springer},
  doi = {10.1007/s00220-016-2706-8}
}

@article{Hoggar1984,
  author = {S. G. Hoggar},
  title = {Parameters of $t$-designs in $\mathbb{F} {P}^{d-1}$},
  journal = {European Journal of Combinatorics},
  volume = {5},
  pages = {29},
  year = {1984},
  doi = {10.1016/S0195-6698(84)80015-3}
}

@article{BannaiHoggar1985,
  author = {E. Bannai and S. G. Hoggar},
  title = {On tight $t$-designs in compact symmetric spaces of rank one},
  journal = {Proceedings of the Japan Academy},
  volume = {61A},
  pages = {78},
  year = {1985},
  nolink = {}
}

@article{Hoggar1989,
  author = {S. G. Hoggar},
  title = {Tight 4 and 5-designs in projective spaces},
  journal = {Graphs and Combinatorics},
  volume = {5},
  pages = {87},
  year = {1989},
  doi ={10.1007/BF01788661}
}

@article{Hoggar1992,
  author = {S. G. Hoggar},
  title = {$t$-designs with general angle set},
  journal = {European Journal of Combinatorics},
  volume = {13},
  pages = {257},
  year = {1992},
  doi = {10.1016/S0195-6698(05)80032-0}
}

@article{czartowski2024quantum,
  title = {Quantum pushforward designs},
  author = {Czartowski, Jakub and \ifmmode \dot{Z}\else \.{Z}\fi{}yczkowski, Karol},
  journal = {Phys. Rev. A},
  volume = {111},
  issue = {3},
  pages = {032433},
  numpages = {15},
  year = {2025},
  month = {Mar},
  publisher = {American Physical Society},
  doi = {10.1103/PhysRevA.111.032433},
  url = {https://link.aps.org/doi/10.1103/PhysRevA.111.032433}
}

@article{Simmonett2022,
  author  = {Simmonett, Andrew C. and Brooks, Bernard R. and Darden, Thomas A.},
  title   = {Efficient and scalable electrostatics via spherical grids and treecode summation},
  journal = {The Journal of Chemical Physics},
  year    = {2025},
  volume  = {162},
  number  = {22},
  pages   = {224101},
  doi     = {10.1063/5.0264934},
  pmid    = {40493039},
  pmcid   = {PMC12158466},
  url     = {https://doi.org/10.1063/5.0264934}
}

@article{iosue2024continuous,
  title     = {Continuous-variable quantum state designs: theory and applications},
  author    = {Iosue, Joseph T and Sharma, Kunal and Gullans, Michael J and Albert, Victor V},
  journal   = {Physical Review X},
  volume    = {14},
  number    = {1},
  pages     = {011013},
  year      = {2024},
  publisher = {APS},
  doi = {10.1103/PhysRevX.14.011013}
}

@article{Markiewicz2023,
  title = {Duality of averaging of quantum states over arbitrary symmetry groups revealing {Schur--Weyl} duality},
  volume = {56},
  ISSN = {1751-8121},
  url = {http://dx.doi.org/10.1088/1751-8121/acf4d5},
  DOI = {10.1088/1751-8121/acf4d5},
  number = {39},
  journal = {Journal of Physics A: Mathematical and Theoretical},
  publisher = {IOP Publishing},
  author = {Markiewicz,  Marcin and Przewocki,  Janusz},
  year = {2023},
  month = sep,
  pages = {395301}
}

@inproceedings{Hallgren2008superpolynomial,
  author    = {Sean Hallgren and Aram W. Harrow},
  editor    = {Luca Aceto and Ivan Damg{\aa}rd and Leslie Ann Goldberg and Magn{\'{u}}s M. Halld{\'{o}}rsson and Anna Ing{\'{o}}lfsd{\'{o}}ttir and Igor Walukiewicz},
  title     = {Superpolynomial Speedups Based on Almost Any Quantum Circuit},
  booktitle = {Automata, Languages and Programming, 35th International Colloquium, {ICALP} 2008, Reykjavik, Iceland, July 7-11, 2008, Proceedings, Part {I:} Track {A:} Algorithms, Automata, Complexity, and Games},
  series    = {Lecture Notes in Computer Science},
  volume    = {5125},
  pages     = {782-795},
  publisher = {Springer},
  year      = {2008},
  url       = {https://doi.org/10.1007/978-3-540-70575-8\_64},
  doi       = {10.1007/978-3-540-70575-8\_64},
  bibsource = {dblp computer science bibliography, https://dblp.org}
}

@article{sawicki2022universality,
  title = {Universality verification for a set of quantum gates},
  author = {Sawicki, Adam and Mattioli, Lorenzo and Zimbor\'as, Zolt\'an},
  journal = {Phys. Rev. A},
  volume = {105},
  issue = {5},
  pages = {052602},
  numpages = {6},
  year = {2022},
  month = {May},
  publisher = {American Physical Society},
  doi = {10.1103/PhysRevA.105.052602},
  url = {https://link.aps.org/doi/10.1103/PhysRevA.105.052602}
}

@article{zyczkowski2001induced,
  title     = {Induced measures in the space of mixed quantum states},
  author    = {{\.Z}yczkowski, Karol and Sommers, Hans-J{\"u}rgen},
  journal   = {Journal of Physics A: Mathematical and General},
  volume    = {34},
  number    = {35},
  pages     = {7111},
  year      = {2001},
  publisher = {IOP Publishing},
  doi = {10.1088/0305-4470/34/35/335}
}

@article{DATTA20122455,
title = {Geometry of the {Welch} bounds},
journal = {Linear Algebra and its Applications},
volume = {437},
number = {10},
pages = {2455-2470},
year = {2012},
issn = {0024-3795},
doi = {https://doi.org/10.1016/j.laa.2012.05.036},
url = {https://www.sciencedirect.com/science/article/pii/S0024379512004405},
author = {S. Datta and S. Howard and D. Cochran}
}

@article{leone2025non1clifford,
  title = {{Non-Clifford Cost of Random Unitaries}},
  author = {Leone, Lorenzo and Oliviero, Salvatore F.E. and Hamma, Alioscia and Eisert, Jens and Bittel, Lennart},
  journal = {PRX Quantum},
  volume = {7},
  issue = {2},
  pages = {020321},
  numpages = {29},
  year = {2026},
  month = {May},
  publisher = {American Physical Society},
  doi = {10.1103/25v1-my1x},
  url = {https://link.aps.org/doi/10.1103/25v1-my1x}
}

@article{Jian2023,
  author       = {Jian, Shao-Kai and Bentsen, Gregory and Swingle, Brian},
  title        = {Linear growth of circuit complexity from Brownian dynamics},
  journal      = {Journal of High Energy Physics},
  year         = {2023},
  volume        = {2023},
  number        = {8},
  pages         = {190},
  doi           = {10.1007/JHEP08(2023)190},
  publisher     = {Springer},
  url           = {https://doi.org/10.1007/JHEP08(2023)190}
}

\end{document}